\documentclass{article}
\usepackage{cmap}
\usepackage[utf8]{inputenc}
\inputencoding{utf8}
\usepackage{amsmath,amssymb,amsthm,graphicx}
\usepackage{mathtools}
\usepackage{authblk}
\usepackage{xcolor}
\usepackage{texdraw}
\usepackage[pdfpagemode=UseOutlines,
bookmarks=true, bookmarksopenlevel=3, bookmarksopen=true,
bookmarksnumbered=true, unicode=true,
pdffitwindow=true,pdfpagelayout=SinglePage,pdfhighlight=/P,
colorlinks=true,linkcolor=teal,citecolor=blue,urlcolor=blue]{hyperref}
\usepackage[margin=1.25in]{geometry}
\DeclareMathOperator{\ArcSinh}{ArcSinh}

\newcommand{\al}{\alpha}
\newcommand{\be}{\beta}
\theoremstyle{definition}
\newtheorem{definition}{Definition}
\theoremstyle{remark}
\newtheorem*{remark}{Remark}

\theoremstyle{plain}
\newtheorem{prop}{Proposition}
\title{multi-quadratic equations associated with KdV-type equations}
\author[]{M.P. Kolesnikov\thanks{kolesnikov.mp@phystech.edu}} 
\affil[]{Center of Integrable Systems, P.G. Demidov Yaroslavl State University, Yaroslavl, Russia}
\date{}
\begin{document}
\newtheorem{theorem}{Theorem}[section]
\hypersetup{hypertexnames=false}
\maketitle
\begin{abstract}
The potential KdV equation admits dpKdV as a full discretization. Similarly, most KdV-type equations admit multi-quadratic quad equations as algebraic condition on B\"acklund transformation. These quad equations also yield three-point discrete symmetries and ensure the commutativity of two symmetries corresponding to different lattice directions. In each case, the quad equation is either an ABS equation or a multi-quadratic equation of the Atkinson--Nieszporski class. For KdV, mKdV, potential Gardner and exponential Calogero--Degasperis, these are the equations $H2^*$, $A1^*$, $Q1^*$ and $A2^*$, respectively, and for the elliptic Calogero--Degasperis equation it is the primary model $Q4^*$.
\end{abstract}
\bigskip
\begin{quotation}
\end{quotation}

\section{Introduction}\label{sec:introduction}
Integrable systems arise in many areas of mathematical and theoretical physics, including hydrodynamics, nonlinear optics, plasma physics, statistical mechanics and field theory. One of their characteristic features is the existence of soliton solutions. Optical solitons are used in fibre-optic communications \cite{Hasegawa_Tappert}, and magnetic flux quanta (fluxons) in long Josephson junctions behave as sine-Gordon solitons \cite{Ustinov}.

The symmetry approach is one of the most effective tools in the theory of integrable systems \cite{MSY_1987,Mikhailov_Shabat_Sokolov,Sokolov_book}. In this approach, integrability is identified with the existence of an infinite hierarchy of generalized symmetries. The approach has led, in particular, to the classification of integrable KdV-type equations \cite{Svinolupov_Sokolov_1982,MSY_1987,Meshkov_Sokolov} and of integrable Volterra-type lattices \cite{Yamilov_1983,Yamilov}.

An important aspect of the theory is the relationship between continuous equations and their semidiscrete and fully discrete counterparts. A Bäcklund transformation can be regarded as a relation between fields at neighbouring sites of a one-dimensional lattice. It is therefore a differential-difference equation, often called a dressing chain \cite{Levi_1981,Shabat_Yamilov_1990,Veselov_Shabat}. Bäcklund transformations produce new solutions from a given one, and their repeated application yields multisoliton solutions. By the Bianchi permutability theorem, two Bäcklund transformations with different parameters commute. The resulting nonlinear superposition formula is an algebraic relation among four solutions \cite{Wahlquist_Estabrook,Rogers_Schief}. For the potential KdV equation, this relation is the discrete potential KdV equation \cite{Hirota_1977,NQCL_1984,Nijhoff_Capel}. Conversely, continuous and semidiscrete equations can be recovered from fully discrete ones by suitable continuum limits. Since discrete equations are algebraic, it is often easier to construct their exact solutions \cite{HJN_book}.

Multidimensional consistency was proposed as an integrability criterion for quad equations in \cite{Nijhoff_2002,Bobenko_Suris_2002}. It means that the equation can be consistently imposed on all faces of a cube, and hence on a lattice of any dimension. A Lax pair and a Bäcklund transformation then follow from the equation itself. Adler, Bobenko and Suris classified the multiaffine quad equations that are consistent around a cube, have the symmetries of the square and have the tetrahedron property \cite{ABS}. This result was later extended to the case without these symmetry assumptions \cite{ABS_2009}. The top equation of the ABS list, $Q_4$, first appeared as the nonlinear superposition principle for the Bäcklund transformation of the Krichever--Novikov equation \cite{Krichever_Novikov,Adler_1998,Nijhoff_2002,Adler_Suris_Q4}. Atkinson and Nieszporski introduced the factorized-discriminant hypothesis and used it to obtain integrable multi-quadratic quad equations \cite{Atkinson}. Further multi-quadratic examples arise from the quasi-classical limit of the star-triangle relation \cite{Kels}. Bäcklund transformations for the ABS equations themselves were studied in \cite{Atkinson_2008}. The generalized symmetries of the ABS equations are three-point differential-difference equations of Volterra type in each lattice direction \cite{Rasin_Hydon,Tongas_Tsoubelis_Xenitidis,LPSY_2008,Tsoubelis_Xenitidis}. These are the discrete symmetries considered in the present paper.

A closely related structure arises in the theory of negative symmetries \cite{Verosky,Adler_2024,Kolesnikov_2025}. These are nonlocal symmetries of a particular form that extend the algebra of generalized symmetries. Reductions with respect to negative symmetries lead to Painlevé-type equations and their multicomponent analogues \cite{AK_2022,AK_2023}. Negative symmetries can also be obtained from compatible differential-difference chains. Given a Bäcklund transformation and a Volterra-type lattice that are compatible, the elimination of the discrete variable yields a negative symmetry \cite{Adler_2024,Kolesnikov_2025}. Bäcklund transformations for integrable evolution equations were studied in \cite{Svinolupov_Sokolov_Yamilov}, and semidiscrete equations admitting third-order symmetries were classified in \cite{Garif_2023}; see also \cite{Garif_2021}.

The present paper studies quad equations associated with the KdV-type equations of the Meshkov--Sokolov list \cite{Meshkov_Sokolov}. The main results are as follows.
\begin{itemize}
  \item We reproduce the Bäcklund transformations obtained in \cite{Garif_2023}. We pay particular attention to the integration constants, since they play the role of the Bäcklund  parameters. The procedure used to obtain these results is described in detail in the Appendix.
  \item For each equation of the Meshkov--Sokolov list that admits a Bäcklund transformation of the required form, we derive the associated multiaffine or multi-quadratic quad equation from the Bianchi permutability diagram. The multiaffine equations are H1, $Q1_{\delta=0}$ and Q4 of the ABS list. We identify the multi-quadratic equations for KdV, mKdV, potential Gardner and  exponential Calogero--Degasperis with the equations $H2^*$, $A1^*$, $Q1^*$ and $A2^*$ of
  \cite{Atkinson}. This gives their 3D consistency and their relation to ABS equations. The multi-quadratic equation for the elliptic Calogero--Degasperis equation with cubic $q$ is the primary model $Q4^*$, up to a Möbius change of variables.
  \item We prove that the quad equations are compatible with the flows of the discrete symmetries, modulo the quad equations and their shifts. We also prove that the flows generated by the two discrete symmetries for the two lattice directions commute.
  \item We give a simple method for deriving three-point discrete symmetries from quad equations. For multiaffine equations, it reproduces the known formula for the symmetries of the ABS equations in terms of the edge biquadratics \cite{Tsoubelis_Xenitidis,Xenitidis_2009}. We extend it to multi-quadratic equations by means of the vertex discriminant. In most cases, the method yields the general three-point symmetry.
\end{itemize}

The paper is organized as follows.

In Section~\ref{sec:preliminaries}, we introduce the notation used throughout the paper. We describe the jet-space formalism and define compatibility between an evolutionary equation and a Bäcklund transformation. Using the potential KdV equation as an example, we show how a quad equation can be derived from Bäcklund transformations arranged in a Bianchi permutability diagram. Here, the quad equation is the discrete potential KdV equation. At the end of the section, we introduce the notation for the lattice vertices used in the subsequent calculations.

In Section~\ref{sec:triplets}, we explain how B\"acklund transformations can be obtained by solving the compatibility conditions on jet spaces. We also describe how these transformations depend on the derivatives for the equations of the Meshkov--Sokolov list. We then introduce compatibility between a Bäcklund transformation and a discrete symmetry. Next, we describe how to prove that a quad equation is compatible with the flow of a discrete symmetry and that the discrete symmetries for the two lattice directions commute. Finally, using the Krichever--Novikov equation as an example, we show how a three-point symmetry can be derived from a quad equation.

In Section~\ref{sec:quads}, we treat the equations of the Meshkov--Sokolov list one by one. For each equation, we present the Bäcklund transformation, derive the associated quad equation and the three-point discrete symmetry, and prove that the KdV-type equation, the B\"acklund transformation and the discrete symmetry are pairwise compatible. We then show that the quad equation is compatible with the flows of the discrete symmetries and that the two flows commute, modulo the quad equation and its shifts.

In Section~\ref{sec:integrability}, we study the integrability of the multi-quadratic equations. We identify the multi-quadratic equations with known ones, write the associated Miura-type maps and Yang--Baxter maps in the present notation, discuss the elliptic case.

The Appendix contains the solution of the determining equations. These equations give the general B\"cklund transformations and the general three-point discrete symmetries.

\section{Preliminaries}\label{sec:preliminaries}

In this section, we introduce the notation, briefly explain the jet space framework, and illustrate how a full discretization of the potential Korteweg de Vries equation can be obtained from the commutativity of two B\"acklund transformations.

We use the following conventions throughout.
\begin{itemize}
  \item The variables $x,t,\tau$ are continuous. The field $u$ also depends on one or two integer lattice indices. We denote its first three $x$-derivatives by $u_{x},u_{xx},u_{xxx}$, that is, $\partial_xu,\partial^2_xu,\partial^3_xu.$
  \item A B\"acklund transformation relates fields at adjacent sites, $u_n\to u_{n+1}$. On a two-dimensional lattice, $\al, \be$ denote the parameters in the horizontal and vertical directions, respectively.
  \item $D,D^{-1}$ denote the total $x$-derivative and its formal inverse. The operator $T$ shifts the lattice index by one.
  \item We write $u_{nm}$ for a field with two lattice indices and $u_{nm,x},u_{nm,xx}\ldots $ for its successive $x$-derivatives. By translational invariance, we consider the equations in the neighborhood of the lattice point $u_{00}=u$.
\end{itemize}

\subsection{Jet variables and evolutionary vector fields}

Consider a differential equation of order $n$ of the form
\begin{equation}
   f(x,u,\partial_xu,\ldots \partial^n _xu)=0
\end{equation}
We consider the infinite jet space, a manifold on which $x,u,\partial_xu,\ldots \partial^n _xu\ldots $ are independent coordinates. An evolutionary equation
\begin{equation}
    u_t=F(x,u,\partial_xu,\ldots \partial^n _xu)
\end{equation}
defines dynamics on this manifold. The total $x$-derivative and the evolutionary derivation associated with the characteristic $F$ are given by
\begin{equation}\label{D_x}
    D_{x}=D=\partial_{x}+u_{x}\partial_{u}+u_{xx}\partial_{u_{x}}+\ldots
\end{equation}
\begin{equation}\label{D_t_1}
    D_{t}=F\partial_{u}+D\left(F\right)\partial_{u_x}+D^2(F)\partial_{u_{xx}}+ \ldots,
\end{equation}
We can also consider a jet space in which u depends on an integer parameter. This leads to the following definition.
\begin{definition}
Let $v=Tu$. The evolution equation
    \begin{equation}\label{KdV_def}
         u_{t}=F\left(x,u,u_x,u_{xx},u_{xxx}\right),
    \end{equation}
and the first-order differential relation
    \begin{equation}\label{dress_def}
        v_x=h_+\left(x,v,u,u_x\right)
    \end{equation}
are called \emph{consistent} if the shift and the two derivations satisfy the commutation relations
    \begin{equation}\label{def_1}
    \left[T,D_{x}\right]=0, \quad  \left[D_t,D_{x}\right]=0,\quad  \left[T,D_{t}\right]=0.
    \end{equation}
The first identity in \eqref{def_1} is understood on differential functions $g(u,u_x,u_{xx}\ldots)$. For equations \eqref{KdV_def} and \eqref{dress_def}, these conditions are equivalent to the following equation:
 \begin{equation}\label{con_1}
    TD_x\left(F\right)=D_t\left(h_+\right).
    \end{equation}
\end{definition}
\subsection{Quad equations and Bianchi permutability}
\begin{figure}[ht]
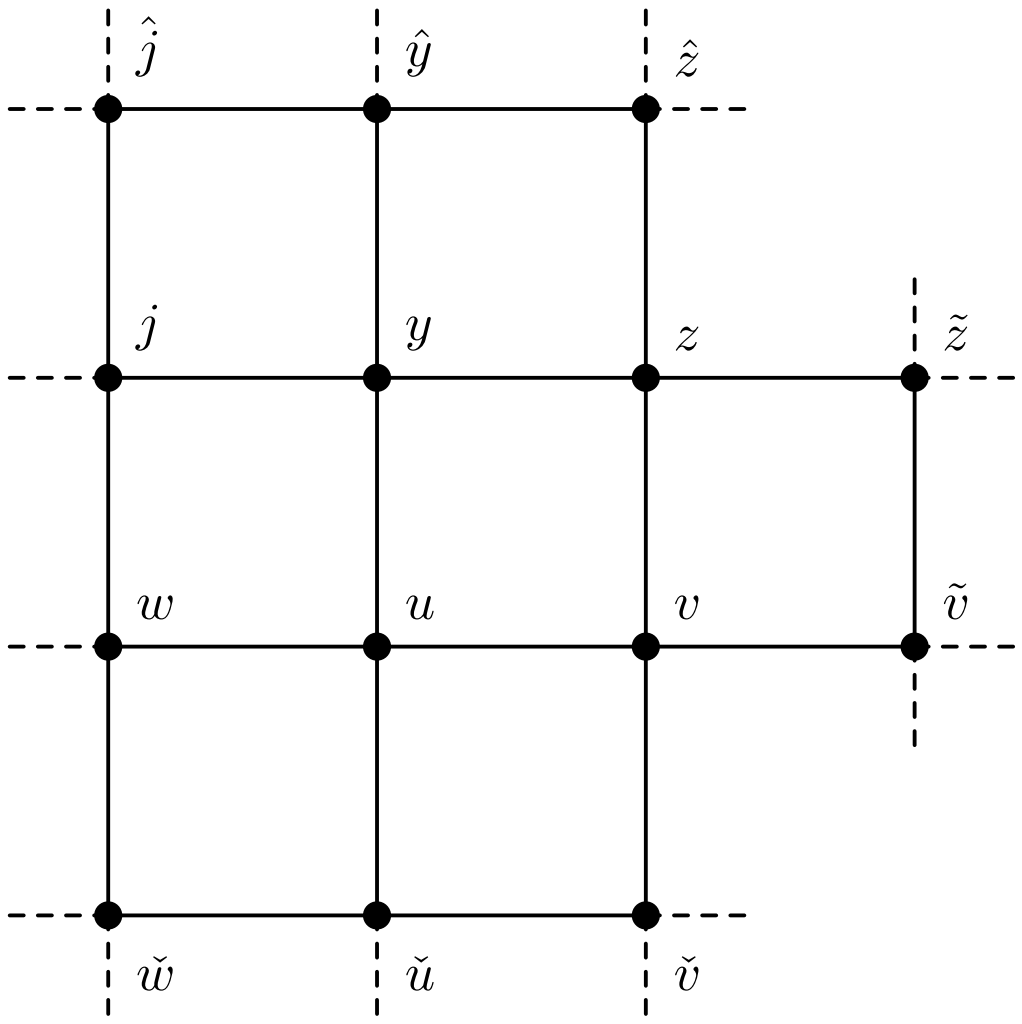

\centering
\centertexdraw{
  \setunitscale .7

  \move (0 0) \lvec (2 0)
  \move (0 1) \lvec (3 1)
  \move (0 2) \lvec (3 2)
  \move (0 3) \lvec (2 3)

  \move (0 0) \lvec (0 3)
  \move (1 0) \lvec (1 3)
  \move (2 0) \lvec (2 3)
  \move (3 1) \lvec (3 2)

  \lpatt (0.05 0.05)

  \move (0 0) \lvec (0 -.4)
  \move (1 0) \lvec (1 -.4)
  \move (2 0) \lvec (2 -.4)
  \move (3 1) \lvec (3 .6)

  \move (0 3) \lvec (0 3.4)
  \move (1 3) \lvec (1 3.4)
  \move (2 3) \lvec (2 3.4)
  \move (3 2) \lvec (3 2.4)

  \move (0 0) \lvec (-.4 0)
  \move (0 1) \lvec (-.4 1)
  \move (0 2) \lvec (-.4 2)
  \move (0 3) \lvec (-.4 3)

  \move (2 0) \lvec (2.4 0)
  \move (3 1) \lvec (3.4 1)
  \move (3 2) \lvec (3.4 2)
  \move (2 3) \lvec (2.4 3)

  \lpatt ()

  \move (0 0) \fcir f:0 r:.05
  \move (1 0) \fcir f:0 r:.05
  \move (2 0) \fcir f:0 r:.05

  \move (0 1) \fcir f:0 r:.05
  \move (1 1) \fcir f:0 r:.05
  \move (2 1) \fcir f:0 r:.05
  \move (3 1) \fcir f:0 r:.05

  \move (0 2) \fcir f:0 r:.05
  \move (1 2) \fcir f:0 r:.05
  \move (2 2) \fcir f:0 r:.05
  \move (3 2) \fcir f:0 r:.05

  \move (0 3) \fcir f:0 r:.05
  \move (1 3) \fcir f:0 r:.05
  \move (2 3) \fcir f:0 r:.05

  \htext (.10 -.28) {$\check{w}$}
  \htext (1.10 -.28) {$\check{u}$}
  \htext (2.10 -.28) {$\check{v}$}

  \htext (.10 1.10) {$w$}
  \htext (1.10 1.10) {$u$}
  \htext (2.10 1.10) {$v$}
  \htext (3.10 1.10) {$\tilde{v}$}

  \htext (.10 2.10) {$j$}
  \htext (1.10 2.10) {$y$}
  \htext (2.10 2.10) {$z$}
  \htext (3.10 2.10) {$\tilde{z}$}

  \htext (.10 3.12) {$\hat{j}$}
  \htext (1.10 3.12) {$\hat{y}$}
  \htext (2.10 3.12) {$\hat{z}$}
}
\caption{Vertex notation on the lattice. The central elementary square has vertices $u,v,y,z$.}
\label{fourteen-vertex-graph}
\end{figure}

Given an evolutionary equation and a B\"acklund transformation, we can, in some cases, construct an algebraic equation for a variable with two discrete parameters. We call this a quad equation \cite{Bobenko_Suris_2002}.
An important source of quad equations is the permutability of B\"acklund transformations. Consider as an example the potential KdV equation
\begin{equation}
u_t=u_{xxx}+6u_x^2.
\end{equation}
Its well-known auto-B\"acklund transformation is
\begin{equation}
\widetilde u_x+u_x=-(\widetilde u-u)^2+\alpha,
\end{equation}
where $\alpha$ is a parameter. For two transformations with distinct parameters $\alpha$ and $\beta$, Bianchi permutability means that the two intermediate solutions admit a common transform. With compatible choices of integration constants, this gives
\begin{equation}
B_\beta\bigl(B_\alpha(u)\bigr)=B_\alpha\bigl(B_\beta(u)\bigr).
\end{equation}
We therefore associate the four solutions with the vertices of an
elementary square and write
\[
u=u_{0,0},\qquad v=u_{1,0},\qquad
y=u_{0,1},\qquad z=u_{1,1},
\]
where
\[
v=B_\alpha(u),\qquad y=B_\beta(u),\qquad
z=B_\beta(v)=B_\alpha(y).
\]
The four edges carry the B\"acklund relations
\begin{equation}
\begin{aligned}
v_x+u_x&=-(v-u)^2+\alpha,\quad y_x+u_x=-(y-u)^2+\beta,\\
z_x+v_x&=-(z-v)^2+\beta,\quad z_x+y_x=-(z-y)^2+\alpha.
\end{aligned}
\end{equation}
Add the two horizontal edge relations and subtract the two vertical ones. Each derivative occurs once with each sign. This gives the algebraic relation
\begin{equation}
0=2(\alpha-\beta)-(v-u)^2+(y-u)^2+(z-v)^2-(z-y)^2.
\end{equation}
After factorization, we obtain the discrete potential KdV equation
\begin{equation}
(z-u)(v-y)-\alpha+\beta=0.
\end{equation}
For $v\ne y$, solving for the fourth vertex gives
\begin{equation}
z=u+\frac{\alpha-\beta}{v-y}.
\end{equation}
This gives the nonlinear superposition formula for the two B\"acklund transformations. More generally, a relation
\begin{equation}
Q(u,v,y,z;\alpha,\beta)=0
\end{equation}
among the four vertex values is called a \emph{quad equation}. Repeating the same construction at each lattice site gives the shifts of the quad equation. Figure~\ref{fourteen-vertex-graph} fixes the vertex notation used below. A tilde denotes one further shift to the right, and a hat or a check denotes a vertical shift up or down, respectively.

\section{Construction of the dressing chain and its discrete symmetries}\label{sec:triplets}
In this section, we describe how equations on jet space can be solved to obtain a B\"acklund transformation for an evolutionary equation. In particular, we derive a condition on B\"acklund transformations for equations in the Meshkov-Sokolov list. We then formulate the compatibility condition for a Volterra chain and a B\"acklund transformation. In the third subsection, we explain how a negative symmetry can be obtained from a B\"acklund transformation and a Volterra-type chain. We use the negative symmetry of potKdV as an example and state its property of continuous 3D consistency. Finally, using the Krichever--Novikov equation as an example, we show how a three-point discrete symmetry can be obtained from a quad equation.

\subsection{B\"acklund transformations on jet space}
We first derive a necessary condition on a B\"acklund transformation. Consider an evolution equation of KdV type with lower-order terms of the form given in the Meshkov-Sokolov list.
\begin{equation}
u_t=u_{xxx}+f_1(u_x)u^2_{xx}+f_2(x,u,u_x)u_{xx}+f_3(x,u,u_x)=F(x,u,u_x,u_{xx},u_{xxx})
\end{equation}
Introduce a lattice index $n\in\mathbb{Z}$. Denoting the next field by $(Tu=v)$, we seek a first-order relation of the form
\begin{equation}\label{Backl-form}
    V=P(x,u,v,u_x,v_x)=0
\end{equation}
According to our definition, the vector field corresponding to the evolutionary equation is
\begin{equation}
    D_t=F\partial_u+D_x(F)\partial_{u_x}+\ldots
\end{equation}
and the consistency condition reduces to the following equation:
\begin{equation}
    D_t(V)=0
\end{equation}
Applying the vector field above yields an equation that is polynomial in some of the jet space variables. Since these variables are independent, each coefficient must vanish. The coefficient of the mixed term $v_{xxx}v_{xx}$ gives
\begin{equation}\label{v3v2}
    2T(f_1)P_{v_x}+\frac{2f_1P_{v_x}^2}{P_{u_x}}-\frac{3P_{u_xu_x}P^2_{v_x}}{P^2_{u_x}}+\frac{6P_{v_x}P_{v_xu_x}}{P_{u_x}}-3P_{v_xv_x}=0
\end{equation}
To illustrate this condition, consider the potential KdV equation
\begin{equation}\label{pkdv-negative}
    u_{t}=u_{xxx}-6u^2_{x}. 
\end{equation}
Its compatible dressing chain is
\begin{equation}\label{dress_pot}
    u_{n+1,x}+u_{n,x}=\left(u_{n+1}-u_{n}\right)^{2}+\alpha\:.
\end{equation}
For potential KdV, the coefficient $f_1$ vanishes, and
\eqref{v3v2} specializes to
\begin{equation}\label{v3v2_pot}
    -\frac{3P_{u_xu_x}P^2_{v_x}}{P^2_{u_x}}+\frac{6P_{v_x}P_{v_xu_x}}{P_{u_x}}-3P_{v_xv_x}=0
\end{equation}
 In the  form $P=v_x-G(x,u,v,u_x)$, equation~\eqref{v3v2_pot} reduces to $G_{u_xu_x}=0$. Thus the transformed derivative at the next site is affine in the derivative at the current site. Only derivatives of $P$ with respect to $u_x$ and $v_x$ occur in \eqref{v3v2}. To obtain an analogous reduction for general $f_1$, seek a relation of the form $P(x,u,v,g(u_x),g(v_x))$. The chain rule gives
{\small
\begin{equation}
    T(2f_1g'(u_x)-3g''(u_x))P_{v_x}+\frac{g'(v_x)^2(2f_1g'(u_x)-3g''(u_x))P_{v_x}^2}{g'(u_x)^2P_{u_x}}-g'(v_x)^2(-\frac{3P_{u_xu_x}P^2_{v_x}}{P^2_{u_x}}+\frac{6P_{v_x}P_{v_xu_x}}{P_{u_x}}-3P_{v_xv_x})=0
\end{equation}
}
\par
Choose $g$ to satisfy $2f_1g'(u_x)-3g''(u_x)=0$. This reduces the determining equation to the affine case above. With $u$ held fixed if $f_1$ depends on it, the first integration gives the following expression, where $C\ne0$:
\[
 g'(p)=C\exp\!\left(\frac{2}{3}\int^p f_1(s)\,ds\right).
\]
For equations in the Meshkov-Sokolov list, $f_1$ takes one of the four forms listed below. We therefore obtain the following conditions on B\"acklund transformations:
\begin{equation}\label{Appcond}
    \begin{array}{llr}
f_1=  0,& v_x = h_3u_x + h_4, & \text{(A)} \\
f_1 =  -\dfrac{3}{4u_1},& v_x^{1/2} = h_3u_x^{1/2} + h_4, \; h_4 \neq 0, & \text{(B)} \\
f_1=  -\dfrac{3}{2u_x},& \log(v_x) = h_3\log(u_x)+h_4, & \text{(C)} \\
f_1=  -\dfrac{3u_x}{2(u_x^2 + Q^2)}, & \ArcSinh{\frac{v_x}{Q}}=h_3\ArcSinh{\frac{u_x}{Q}}+h_4 & \text{(D)}
\end{array}
\end{equation}
where the functions $h_i$ do not depend on derivatives of the variables. 
\begin{remark}
    We assume the absence of pseudoconstants, i.e., that the only solutions of the  equations $D_x(c)=0 \quad (T-1)c=0$   are constants.
\end{remark}

\subsection{Volterra-type chains}
A B\"acklund transformation can be used to construct a differential-difference chain. We now define the consistency condition for a B\"acklund transformation and a three-point symmetry.
\begin{definition}
Consider the three-point lattice flow
\begin{equation}\label{tau-flow}
    u_\tau=g(u,v,w),
\end{equation}
with notations  $v=Tu,u=Tw$. See Figure \ref{fourteen-vertex-graph}.

and a dressing chain
\[
v_x=h_+(x,v,u,u_x),
\]
These equations are called \emph{consistent} if the two total derivatives and the shift obey
\[
 \left[D_{\tau},D_{x}\right]=0,\quad \left[D_{x},T\right]=0,\quad \left[D_{\tau},T\right]=0
\] 
These conditions are equivalent to the single equation
\begin{equation}\label{condition_2}
    D_xT\left(g\right)=D_{z}\left(h_+\right).
\end{equation}
\end{definition}
\begin{remark}
    We will consider discrete symmetries only form \eqref{tau-flow}. If we write down "general symmetry"  it means general symmetry of form \eqref{tau-flow}.
\end{remark}
For the pot-KdV equation, solving the consistency condition \eqref{condition_2} yields the following three-point symmetry:
\begin{equation}\label{KdV-tau}
    u_\tau=\frac{1}{v-w}
\end{equation}
\subsection{Invariance of the quad equation and commutativity of the lattice flows}\label{sec-flows}
The potential KdV equation \eqref{pkdv-negative}, its B\"acklund transformation \eqref{dress_pot}, and the three-point lattice flow \eqref{KdV-tau} are pairwise consistent. Eliminating the lattice shifts between the last two relations gives the negative symmetry \cite{Kolesnikov_2025}
$$u_{xx\tau}=\frac{u_{x\tau}^{2}-\beta^{2}}{2u_{\tau}}+2\left(2u_{x}-\alpha\right)u_{\tau}. $$
\begin{prop}
Consider the two negative symmetries
$$u_{xx\tau_i}=\frac{u_{x\tau_i}^{2}-\beta^{2}}{2u_{\tau_i}}+2\left(2u_{x}-\alpha_i\right)u_{\tau_i} $$
They are three-dimensionally compatible. Their mixed derivative is determined by the
additional equation
   \begin{equation}\label{neg_sup}
       u_{\tau_{2}\tau_{1}}=\frac{1}{2\left(\alpha_{1}-\alpha_{2}\right)}\left(u_{\tau_{1}}u_{x\tau_{2}}-u_{\tau_{2}}u_{x\tau_{1}}\right).
   \end{equation} 
\end{prop}
This means that the two negative symmetries are consistent by virtue of equation \eqref{neg_sup}. The definition of 3D compatibility in the continuous case can be found in the article \cite{Kolesnikov_2025}. For other equations in the Meshkov--Sokolov list, eliminating the lattice shifts can be considerably more involved. We therefore retain the discrete variable. In this description, the quad equation plays the role of the mixed relation \eqref{neg_sup}. Two compatibility statements are checked: first, that each lattice flow preserves the quad equation; second, that the horizontal and vertical flows commute on its solutions.

\begin{figure}[ht]
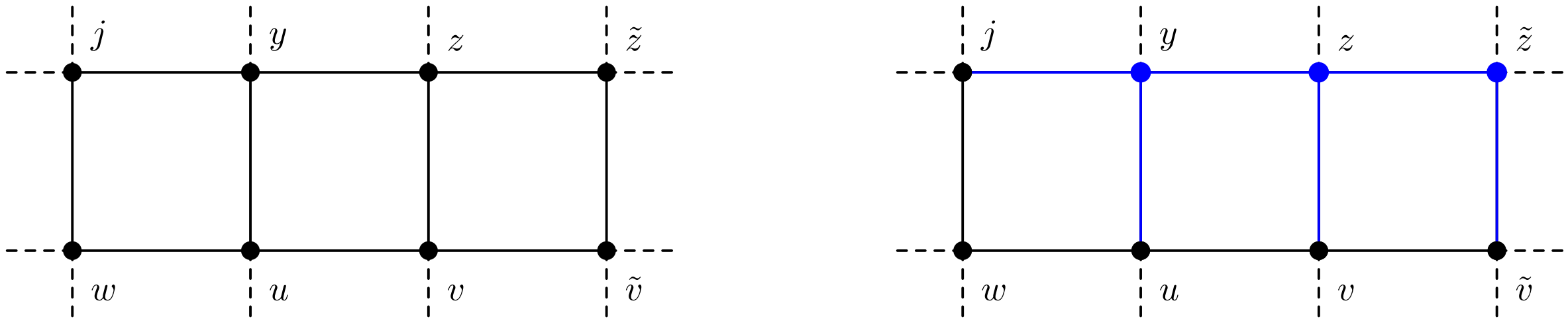

\centering

\def\DrawGrid{%
  \move (0 0) \lvec (3 0)
  \move (0 1) \lvec (3 1)

  \move (0 0) \lvec (0 1)
  \move (1 0) \lvec (1 1)
  \move (2 0) \lvec (2 1)
  \move (3 0) \lvec (3 1)

  \lpatt (0.05 0.05)

  \move (0 0) \lvec (0 -.4)
  \move (1 0) \lvec (1 -.4)
  \move (2 0) \lvec (2 -.4)
  \move (3 0) \lvec (3 -.4)

  \move (0 1) \lvec (0 1.4)
  \move (1 1) \lvec (1 1.4)
  \move (2 1) \lvec (2 1.4)
  \move (3 1) \lvec (3 1.4)

  \move (0 0) \lvec (-.4 0)
  \move (0 1) \lvec (-.4 1)
  \move (3 0) \lvec (3.4 0)
  \move (3 1) \lvec (3.4 1)

  \lpatt ()

  \move (0 0) \fcir f:0 r:.05
  \move (1 0) \fcir f:0 r:.05
  \move (2 0) \fcir f:0 r:.05
  \move (3 0) \fcir f:0 r:.05

  \move (0 1) \fcir f:0 r:.05
  \move (1 1) \fcir f:0 r:.05
  \move (2 1) \fcir f:0 r:.05
  \move (3 1) \fcir f:0 r:.05

  \htext (.10 -.28) {$w$}
  \htext (1.10 -.28) {$u$}
  \htext (2.10 -.28) {$v$}
  \htext (3.10 -.28) {$\tilde{v}$}

  \htext (.10 1.12) {$j$}
  \htext (1.10 1.12) {$y$}
  \htext (2.10 1.12) {$z$}
  \htext (3.10 1.12) {$\tilde{z}$}
}

\def\BlueDot{%
  \writeps{
    gsave
    0 0 1 setrgbcolor
    currentpoint newpath 12 0 360 arc fill
    grestore
  }
}

\centertexdraw{
  \setunitscale .7

  \DrawGrid

  \move (5 0)
  \bsegment

    \DrawGrid

    \writeps{stroke 0 0 1 setrgbcolor}

    \move (0 1) \lvec (3 1)
    \move (1 0) \lvec (1 1)
    \move (2 0) \lvec (2 1)
    \move (3 0) \lvec (3 1)

    \writeps{stroke 0 0 0 setrgbcolor}

    \move (0 1) \fcir f:0 r:.05
    \move (1 0) \fcir f:0 r:.05
    \move (2 0) \fcir f:0 r:.05
    \move (3 0) \fcir f:0 r:.05

    \move (1 1) \BlueDot
    \move (2 1) \BlueDot
    \move (3 1) \BlueDot

  \esegment
}

\caption{The eight vertices entering the horizontal derivative of the central quad equation. The right panel marks the three upper vertices eliminated using consecutive squares.}
\label{quad-graphs}
\end{figure}

Let $\tau$ be the time of the horizontal three-point flow. The chain rule gives its action on the central relation as
\[
 D_\tau Q=Q_u u_\tau+Q_v v_\tau+Q_y y_\tau+Q_z z_\tau.
\]
At each vertex, the characteristic uses its two horizontal neighbours. The union of these sites consists of the eight vertices in Figure~\ref{quad-graphs}. Starting with the left square, solve three consecutive copies of the quad equation for $y$, $z$, and $\tilde z$, respectively.  The reduced derivative vanishes in the cases considered below. This is the meaning of invariance modulo the quad equation and its shifts.

\begin{figure}[ht]
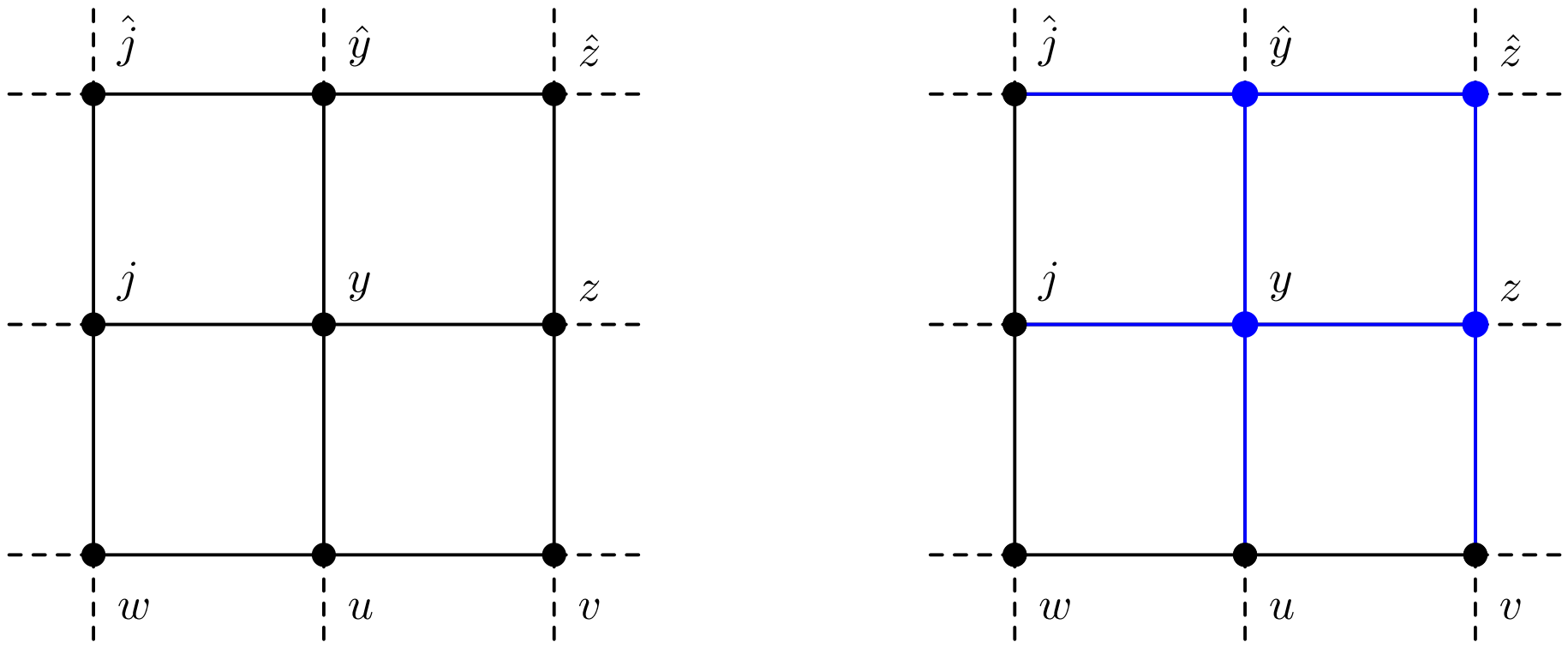

\centering

\def\DrawGrid{%
  \move (0 0) \lvec (2 0)
  \move (0 1) \lvec (2 1)
  \move (0 2) \lvec (2 2)

  \move (0 0) \lvec (0 2)
  \move (1 0) \lvec (1 2)
  \move (2 0) \lvec (2 2)

  \lpatt (0.05 0.05)

  \move (0 0) \lvec (0 -.4)
  \move (1 0) \lvec (1 -.4)
  \move (2 0) \lvec (2 -.4)

  \move (0 2) \lvec (0 2.4)
  \move (1 2) \lvec (1 2.4)
  \move (2 2) \lvec (2 2.4)

  \move (0 0) \lvec (-.4 0)
  \move (0 1) \lvec (-.4 1)
  \move (0 2) \lvec (-.4 2)

  \move (2 0) \lvec (2.4 0)
  \move (2 1) \lvec (2.4 1)
  \move (2 2) \lvec (2.4 2)

  \lpatt ()

  \move (0 0) \fcir f:0 r:.05
  \move (1 0) \fcir f:0 r:.05
  \move (2 0) \fcir f:0 r:.05

  \move (0 1) \fcir f:0 r:.05
  \move (1 1) \fcir f:0 r:.05
  \move (2 1) \fcir f:0 r:.05

  \move (0 2) \fcir f:0 r:.05
  \move (1 2) \fcir f:0 r:.05
  \move (2 2) \fcir f:0 r:.05

  \htext (.10 -.28) {$w$}
  \htext (1.10 -.28) {$u$}
  \htext (2.10 -.28) {$v$}

  \htext (.10 1.10) {$j$}
  \htext (1.10 1.10) {$y$}
  \htext (2.10 1.10) {$z$}

  \htext (.10 2.12) {$\hat{j}$}
  \htext (1.10 2.12) {$\hat{y}$}
  \htext (2.10 2.12) {$\hat{z}$}
}

\def\BlueDot{%
  \writeps{
    gsave
    0 0 1 setrgbcolor
    currentpoint newpath 12 0 360 arc fill
    grestore
  }
}

\centertexdraw{
  \setunitscale .7

  \DrawGrid

  \move (4 0)
  \bsegment

    \DrawGrid

    \writeps{stroke 0 0 1 setrgbcolor}

    \move (0 1) \lvec (2 1)
    \move (0 2) \lvec (2 2)

    \move (1 0) \lvec (1 2)
    \move (2 0) \lvec (2 2)

    \writeps{stroke 0 0 0 setrgbcolor}

    \move (0 0) \fcir f:0 r:.05
    \move (1 0) \fcir f:0 r:.05
    \move (2 0) \fcir f:0 r:.05
    \move (0 1) \fcir f:0 r:.05
    \move (0 2) \fcir f:0 r:.05

    \move (1 1) \BlueDot
    \move (2 1) \BlueDot
    \move (1 2) \BlueDot
    \move (2 2) \BlueDot

  \esegment
}

\caption{The nine vertices used in the commutator calculation. The right panel highlights the upper-right square eliminated and the solid edges incident on its vertices.}
\label{nine-vertex-graphs}
\end{figure}

For commutativity, let $D_{\tau_1}$ and $D_{\tau_2}$ denote the horizontal and vertical flows. We test $(D_{\tau_1}D_{\tau_2}-D_{\tau_2}D_{\tau_1}\bigr)u=0$ using the three-by-three array in Figure~\ref{nine-vertex-graphs}. Each mixed derivative uses the neighbouring sites in both directions, which accounts for the nine vertex values.

If eliminating the highlighted vertex values by the quad relations reduces the commutator to an identity in the remaining data, then the $\tau$ flows commute modulo the quad equation. In some cases, root expressions can be regarded as variables on the edges. In these cases, we also need to eliminate the roots on the blue edges.

\subsection{Recovering a discrete symmetry from a quad polynomial}

The Krichever--Novikov case illustrates how the nonlinear superposition formula determines a three-point characteristic. Its multiaffine quad polynomial has the form
\begin{equation}\label{eq:quad-kn}\begin{aligned}
 Q_{KN}={}&a_1 u v y z+a_2 (u v y+u v z+u y z+v y z)+a_3 (u v+y z)\\ &+a_4 (u y+v z)+a_5 (u z+v y)+a_6 (u+v+y+z)+a_7=0.
\end{aligned}\end{equation}
The horizontal discrete symmetry has the following expression in terms of an edge biquadratic.
\begin{equation}\label{sym_from_quad}
     u_{\tau}=\frac{r(u,v)}{v-w}-\frac12\partial_v r(u,v),
\end{equation}
where the edge biquadratic polynomial is defined by
\begin{equation}
    r(u,v)=\partial_y Q_{KN}\,\partial_z Q_{KN}-Q_{KN}\,\partial_z\partial_y Q_{KN}.
\end{equation}
For several equations in the Meshkov--Sokolov list, the selected quad polynomial is quadratic in each vertex. Its discriminant in one vertex supplies the edge factors needed for a three-point symmetry.

For example, the $KdV$ equation gives the polynomial
\[
\begin{aligned}
Q_{KdV}\coloneqq&(u-z)^2(v-y)^2+(\al-\be)^2\bigl[(u-z)^2+(v-y)^2\bigr]\\
&-2(\al-\be)(u+v+y+z+\al+\be)(u-z)(v-y).
\end{aligned}
\]
The corresponding symmetry has the form
\begin{equation}
    u_\tau=\frac{R(v,u,w)-r(u,v)r(u,w)}{v-w}\quad R(v,u,w)=r(u,v)^2-\frac{1}{2}\partial _v(r(u,v)^2)(v-w)
\end{equation}
Here $r$ and $s$ denote edge roots. For the KdV equation displayed above, they are $r(u,v)=(u+v+\al)^\frac12$ and $s(v,z)=(v+z+\be)^\frac12$. The symmetry and the edge polynomial are related by the discriminant identity
\[
\left(\partial_y Q_{\mathrm{KdV}}\right)^2-2Q_{\mathrm{KdV}}\,\partial_y^2Q_{\mathrm{KdV}}=16(\al-\be)^2(u-z)^2r(u,v)^2s(v,z)^2.
\]

Thus, the factorization of a vertex discriminant provides the edge data for the discrete symmetry without solving a second system of determining equations. In the individual examples below, we specify when $r$ and $s$ instead denote chosen square roots of these edge polynomials.

\section{Quad equations}\label{sec:quads}
For each equation considered below, we first give the B\"acklund transformation. Eliminating the $x$-derivatives from the Bianchi diagram gives a quad equation. In some cases, the roots must be eliminated by squaring to obtain a multi-quadratic quad equation. We then derive a three-point discrete symmetry from the quad equation. We also prove the compatibility of the quad equation with the discrete symmetry, as well as the commutativity of the two discrete symmetries on the lattice.

\subsection{Potential KdV}\label{sec:quad-pkdv}

Consider the potential Korteweg de Vries equation (pot-KdV)
\begin{equation}\label{pKdV}
    u_t=u_{xxx}+6u_x^2,
\end{equation}
Pot-KdV has the following B\"acklund transformation:
\begin{equation}\label{B-pKdV}
    v_x+u_x=-(u-v)^2-k (u-v)+\al
\end{equation}
\begin{prop}
The pot-KdV equation \eqref{pKdV} is consistent with the B\"acklund transformation \eqref{B-pKdV}. Namely,  $\left[D_t,D_{x}\right]=0,$ where $D_x$ is evaluated using the dressing relation and its shifts.
\end{prop}
The nonautonomous transformation $\tilde{u}_n=u_n-\frac12kn$ removes the parameter $k$. We therefore take $k=0$. Eliminating the derivatives from the four edge relations gives the discrete potential KdV equation
\begin{equation}
Q_{pot}\coloneqq-2(z-u)(v-y)+2(\al-\be)=0.
\end{equation}
The edge biquadratic of this multiaffine polynomial is constant. After rescaling the flow parameter, formula $\ref{sym_from_quad}$ yields the discrete symmetry $u_\tau=\frac{1}{v-w}$. The determining equations in the appendix yield the same three-point symmetry with additional constant that corresponds to change of variables $u\to u+\tau c$.

By the explicit substitutions described in Section \ref{sec-flows} we obtain the following theorem
\begin{theorem}\label{thm:pkdv}
The potential KdV equation~\eqref{pKdV}, its B\"acklund transformation~\eqref{B-pKdV}, and the three-point symmetry $u_\tau=\frac{1}{v-w}$ are pairwise consistent. The quad equation is consistent with the discrete symmetry, and the horizontal and vertical $\tau$ flows commute on its solutions. Equivalently, the following identities hold modulo the quad equation and its shifts:
\[
\frac{d}{d\tau}\left[ Q_{pot}\right]=0,\quad \bigl(D_{\tau_1}D_{\tau_2}-D_{\tau_2}D_{\tau_1}\bigr)u=0\quad  mod(Q_{pot})\quad   \text{and its shifts}.
\]
\end{theorem}

\subsection{Korteweg de Vries equation}\label{sec:quad-kdv}
Consider the Korteweg de Vries (KdV) equation
\begin{equation}\label{KdV}
    u_t=u_{xxx}-3uu_x.
\end{equation}
Solving the consistency condition yields the following B\"acklund transformation:
\begin{equation}\label{B-KdV}
    v_x+ u_x= (v-u) \sqrt{ v+u+\al}.
\end{equation}
\begin{prop}
The KdV equation \eqref{KdV} is consistent with the B\"acklund transformation \eqref{B-KdV}. Namely,
$\left[D_t,D_{x}\right]=0,$ with $D_x$ evaluated using the dressing relation and its shifts.
\end{prop}
Eliminating the derivatives on a Bianchi square gives a relation among four edge radicals.

\begin{equation}
\begin{aligned}
&(v-u)\sqrt{v+u+\al}+(z-y)\sqrt{z+y+\al}-\\
 &\quad -(y-u)\sqrt{y+u+\be}-(z-v)\sqrt{z+v+\be}=0.
\end{aligned}
\end{equation}
Set $a^2=A=(v-u)^2(u+v+\al)$, $b^2=B=(z-y)^2(y+z+\al)$, $c^2=C=(y-u)^2(u+y+\be)$, and $d^2D=(z-v)^2(v+z+\be)$. From $a+b=c+d$, we obtain
\[
 A+B-C-D=2(cd-ab),\qquad
(A+B-C-D)^2-4AB-4CD=-8abcd.
\]
Squaring the second identity gives the polynomial 
\begin{equation}\label{elim}
P=\bigl[(A+B-C-D)^2-4AB-4CD\bigr]^2-64ABCD=0.
\end{equation}
The polynomial $P$ factors as $P=F Q_{KdV}$, where $F$ is a polynomial of degree eight and the remaining factor is the multi-quadratic polynomial $Q_{KdV}$ of degree four, given by

\begin{equation}\label{Q_KdV}
    \begin{aligned}
Q_{KdV}\coloneqq&(u-z)^2(v-y)^2+(\al-\be)^2\bigl[(u-z)^2+(v-y)^2\bigr]\\
&-2(\al-\be)(u+v+y+z+\al+\be)(u-z)(v-y).
\end{aligned}
\end{equation}
The discriminant with respect to $y$ factors into polynomials on the two edges adjacent to $v$.
\[
\left(\partial_y Q_{\mathrm{KdV}}\right)^2
-2Q_{\mathrm{KdV}}\,\partial_y^2Q_{\mathrm{KdV}}
=16(\al-\be)^2(u-z)^2(u+v+\al)(v+z+\be).
\]
Put $r=\sqrt{u+v+\al}$, $s=\sqrt{u+y+\be}$. In the following characteristic, $r(u,v)$ denotes the horizontal root, and $R$ is the polarization of its square in the neighboring vertex.
\begin{equation}\label{t-KdV}
    u_\tau=\frac{R(v,u,w)-r(u,v)r(u,w)}{v-w}\quad R(v,u,w)=r(u,v)^2-\frac{1}{2}\partial _v(r(u,v)^2)(v-w)
\end{equation}
Solving $Q_{KdV}$ for $z$ gives the following nonlinear superposition principle for two B\"acklund transformations 
\begin{equation}
     z=u-\frac{(\al-\be)(v-y)}{(\sqrt{u+v+\al }+\sqrt{u+y+\be})^2}
\end{equation}
Substitution of $z$ into the opposite radicands turns them into perfect squares. We therefore obtain the following rational map on the vertex and edge variables:
\begin{equation}\label{map-KdV}
    z=u-\frac{(\al-\be)(v-y)}{(r+s)^2},r_z=s+\frac{(\al-\be)}{(r+s)},s_z=r-\frac{(\al-\be)}{(r+s)}.
\end{equation}
\begin{remark}
    The map on the roots $(r,s)\to(r_z,s_z)$ is an example of a Yang–Baxter map.
\end{remark}
By the explicit substitutions described in Section \ref{sec-flows} we obtain the following theorem
\begin{theorem}\label{thm:kdv}
The KdV equation~\eqref{KdV}, the B\"acklund transformation~\eqref{B-KdV}, and the three-point symmetry~\eqref{t-KdV} are pairwise consistent. The multi-quadratic quad equation \eqref{Q_KdV} is consistent with the $\tau$ flow. The two flows commute modulo the map~\eqref{map-KdV} and its shifts. In formulas,
\[
\frac{d}{d\tau}\left[  z-u+\frac{(\al-\be)(v-y)}{(r+s)^2}\right]=0,\quad \bigl[D_{\tau_1},D_{\tau_2}\bigr]u=0\quad  mod(\text{\eqref{map-KdV}})\quad   \text{and its shifts}.
\]
\end{theorem}

\subsection{Modified KdV}\label{sec:quad-mkdv}
Consider the modified KdV (mKdV) equation
\begin{equation}\label{mKdV}
    u_t=u_{xxx}-6u^2u_x
\end{equation}
Solving the consistency condition yields the following B\"acklund transformation:
\begin{equation}\label{B-mKdV}
    v_x\pm u_x= (u\mp v) \sqrt{ (v\pm u)^2+\alpha}
\end{equation}
The transformations with different signs related  by the transformation $u_n\to(-1)^nu_n$
\begin{prop}
The modified KdV equation \eqref{mKdV} is consistent with the B\"acklund transformation \eqref{B-mKdV}. Namely, $\left[D_t,D_{x}\right]=0,$ with $D_x$ evaluated using the B\"acklund transformation and its shifts.
\end{prop}
Eliminating the derivatives from the four B\"acklund transformations on a Bianchi square gives
\begin{equation}
\begin{aligned}
 &(v-u)\sqrt{(v+u)^2+\al}+(z-y)\sqrt{(z+y)^2+\al}\\
 &\quad -(y-u)\sqrt{(y+u)^2+\be}-(z-v)\sqrt{(z+v)^2+\be}=0.
\end{aligned}
\end{equation}
As in the KdV case, denote the squares of the four edge terms by
\begin{align*}
A&=(v-u)^2((v+u)^2+\al),& B&=(z-y)^2((z+y)^2+\al),\\
C&=(y-u)^2((y+u)^2+\be),& D&=(z-v)^2((z+v)^2+\be).
\end{align*}
The same radical elimination gives
\begin{equation}\label{eq:elim-mkdv}
P=\bigl[(A+B-C-D)^2-4AB-4CD\bigr]^2-64ABCD=0.
\end{equation}
To describe its factors, introduce the linear combinations
\[
S=u+v+y+z,\quad X=u-v+y-z,\quad Y=u+v-y-z,\quad 
\]
The polynomial $P$ has two factors. One has degree six in each vertex, and the other has degreetwo. Write
\[
P=Q_{mKdV}G,
\]
with the multi-quadratic component
\[
Q_{mKdV}\coloneqq(\al X^2-\be Y^2)[4(\al-\be)-X^2+Y^2]+(\al-\be)S^2(X^2-Y^2)=0.
\]
Writing $Q=Q_{mKdV}$, we obtain the vertex discriminant
\[
Q_y^2-2QQ_{yy}
=
256(\al-\be)^2(u-z)^2
\bigl((u+v)^2+\al\bigr)
\bigl((v+z)^2+\be\bigr).
\]

As in the KdV case, the edge polynomials yield the following discrete symmetry:
\begin{equation}\label{t-mKdV}
    u_\tau=\frac{R(v,u,w)-r(u,v)r(u,w)}{v-w},\quad R(v,u,w)=r(u,v)^2-\frac{1}{2}\partial_v(r(u,v)^2)(v-w),
\end{equation}
where $r=((u+v)^2+\al)^\frac12$. This yields the general three-point symmetry for modified KdV.  Solving $Q_{mKdV}$ for $z$ and choosing one sign gives the superposition formula

\begin{equation}\label{n-mKdV}
    z=u-\frac{\frac 12(\al-\be)(v-y)}{(u+v)(u+y)+\frac12\al+\frac12\be-\sqrt{((u+v)^2+\al)((u+y)^2+\be)}}.
\end{equation}

Introduce the roots on the four edges by
\[
r=((u+v)^2+\al)^\frac12,\quad s=((u+y)^2+\be)^\frac12,\quad r_z=((y+z)^2+\al)^\frac{1}{2},\quad s_z=((v+z)^2+\be)^\frac12.
\]
Set $p=u+v$, $t=u+y$, $D=pt+\frac12\al+\frac12\be-rs$. The vertex and the two roots then transform together according to
\begin{equation}\label{map-mKdV}
    \begin{aligned}
z&=u-\frac{\frac{1}{2}(\al-\be)(v-y)}D,\\
r_z&=-s+\frac{(\al-\be)(r-s)}{2D},\\
s_z&=-r+\frac{(\al-\be)(r-s)}{2D}.
\end{aligned}
\end{equation}
By the explicit substitutions described in Section \ref{sec-flows} we obtain the following theorem
\begin{theorem}\label{thm:mkdv}
The modified KdV equation~\eqref{mKdV}, the B\"acklund transformation~\eqref{B-mKdV}, and its three-point symmetry~\eqref{t-mKdV} are pairwise consistent. Each flow preserves the superposition formula~\eqref{n-mKdV}, and the flows in the two lattice directions commute by virtue of~\eqref{n-mKdV}. In formulas,
\[
\frac{d}{d\tau}\left[z-u+ \frac{\frac12(a-b)(v-y)}{(u+v)(u+y)+\frac12(\al+\be)-rs}\right]=0.\quad mod(\text{\eqref{map-mKdV}})\quad   \text{and its shifts}.
\]

\[
 \bigl[D_{\tau_1},D_{\tau_2}\bigr]u=0\quad  mod(\text{\eqref{map-mKdV}})\quad   \text{and its shifts}.
\]

\end{theorem}

\subsection{The exponential Calogero--Degasperis equation}\label{sec:quad-exp}

Consider the exponential Calogero--Degasperis (expCD) equation in the form
\begin{equation}\label{expCD}
    u_t = u_{xxx} - \frac{1}{2}u_x^3 + (c_1e^{2u} + c_2e^{-2u})u_x.
\end{equation}
Solving the consistency condition yields two B\"acklund transformations. The first involves the sum of the derivatives:
\begin{equation}\label{B-expCD1}
    v_x+u_x=(e^{-v}+e^{-u})\sqrt{c_1e^{2(v+u)}+2\al e^{v+u}+c_2}
\end{equation}
The second involves their difference:
\begin{equation}\label{B-expCD2}
    v_x-u_x=\left(\sqrt{c_1}+\sqrt{c_2}\,e^{-v-u}\right)\sqrt{e^{2v}+2 \al e^{v+u}+e^{2u}}
\end{equation}

In addition to the transformations given above, the equation admits two further transformations with the opposite sign, which can be obtained by the substitution $u_n\to u_n+in\pi$, where $i^2=-1$.
\begin{prop}
The exponential Calogero--Degasperis equation~\eqref{expCD} is consistent with the B\"acklund transformations \eqref{B-expCD1} and
\eqref{B-expCD2}. In each case, $\left[D_t,D_{x}\right]=0, $ where $D_x$ is evaluated using the corresponding B\"acklund transformation.
\end{prop}

When $c_1=c_2=0$, the equation also admits the two-parameter transformation
\begin{equation}\label{B-expCD3}
    v_x+ u_x=\al_1e^{(v+u)/2}+\al_2e^{-(v+ u)/2}.
\end{equation}
And additional by the substitution $u_n\to(-1)^nu_n$
\begin{prop}
The exponential Calogero--Degasperis equation \eqref{expCD} with coefficients $c_1=c_2=0$ is consistent with \eqref{B-expCD3}. Namely, $\left[D_t,D_{x}\right]=0, $ with $D_x$
evaluated using \eqref{B-expCD3}.
\end{prop}

We begin with the sum relation~\eqref{B-expCD1}. Placing it on all four edges of the square
and eliminating the derivatives gives

\begin{equation}
\begin{aligned}
 &(e^{-v}+e^{-u})
 \sqrt{c_1e^{2(v+u)}+2\al e^{v+u}+c_2}+(e^{-z}+e^{-y})\sqrt{c_1e^{2(z+y)}+2\al e^{z+y}+c_2}\\
 &\quad -(e^{-y}+ e^{-u})\sqrt{c_1e^{2(y+u)}+2 \be e^{y+u}+c_2} -(e^{-z}+e^{-v})\sqrt{c_1e^{2(z+v)}+2 \be e^{z+v}+c_2}=0.
\end{aligned}
\end{equation}

Introduce exponential vertex coordinates 
\[
U=e^{u},\quad V=e^{v},\quad Y=e^{y},\quad Z=e^{z}.
\]
Clearing denominators and eliminating the radicals gives the factorization
\[
P=(UVZY)^4Q_{exCD}G.
\]

Here $P$ is the resulting polynomial, and $G$ is the complementary factor of degree six in each variable. The multi-quadratic quad equation is
\begin{equation}\label{Q-expCD}
    \begin{aligned}
Q_{exCD}\coloneqq&[\al(UY+VZ)-\be(UV+YZ)]^2-4UVYZ(\al-\be)^2\\
&-c_1c_2(U-Z)^2(V-Y)^2-2(\al-\be)(c_1UVYZ+c_2)(U-Z)(V-Y)=0.
\end{aligned}
\end{equation}

For this component, write $Q=Q_{exCD}$ and differentiate with respect to the exponential vertex $Y$. Its discriminant factors as
\[
(\partial_Y Q)^2-2Q\,\partial_Y^2Q
={}4(\al-\be)^2(U-Z)^2\bigl(c_1U^2V^2+2 \al UV+c_2\bigr)\bigl(c_1V^2Z^2+2 \be VZ+c_2\bigr).
\]

The three-point characteristic associated with the horizontal biquadratic polynomial is written as

\begin{equation}\label{t-expCD}
    U_\tau=\frac{R(V,U,W)-r(U,V)r(U,W)}{V-W}\quad R(V,U,W)=r(U,V)^2-\frac{1}{2}\partial _V(r(U,V)^2)(V-W)
\end{equation}
Where $r(U,V)=(c_1U^2V^2+2 \al UV+c_2\bigr)^\frac12$, and in original coordinates:
\begin{align*}
r&=(c_1e^{2(u+v)}+2 \al e^{u+v}+c_2)^\frac12,s=(c_1e^{2(u+y)}+2 \be e^{u+y}+c_2)^\frac12.
\end{align*}
Solving the quad equation \eqref{Q-expCD} for $e^z$ gives
\begin{equation}\label{n-expCD}
    e^{z}=e^{u}
\frac{(c_1e^{2u+v+y}+c_2)+\al e^{u+y}+\be e^{u+v}\ + rs}
{(c_1e^{2u+v+y}+c_2)+\al e^{u+v}+\be e^{u+y}\ + rs}
\end{equation}
For the induced transformation of the radicals on adjacent edges, denote the edge products and the denominator by
\[
p=\exp(u+v),q=\exp(u+y)\quad D=(c_1pq+c_2)+\al p+\be q+rs.
\]
Substitution into the upper horizontal radicand gives
\[
\begin{aligned}
c_1e^{2(z+y)}+2\al e^{z+y}+c_2=\left[\frac{r(c_1q^2+(\al+\be)q+c_2)
+s((c_1pq+c_2)+\al(p+q))}{D}\right]^2.
\end{aligned}
\]

The corresponding identity on the right vertical edge is
\[
\begin{aligned}
c_1e^{2(z+v)}+2 \be e^{z+v}+c_2=\left[\frac{s(c_1p^2+(\al+\be)p+c_2)+r((c_1pq+c_2)+\be(p+q))}{D}\right]^2.
\end{aligned}
\]
By the explicit substitutions described in Section \ref{sec-flows} we obtain the following theorem
\begin{theorem}\label{thm:expcd}
The exponential CD equation~\eqref{expCD}, the B\"acklund transformation \eqref{B-expCD1}, and its three-point symmetry~\eqref{t-expCD} are pairwise consistent. The nonlinear superposition principle~\eqref{n-expCD} is consistent with the discrete symmetry, and the two discrete symmetries commute by virtue of~\eqref{n-expCD}. Thus,
\[
\frac{d}{d\tau}\left[ e^{z}-e^{u}
\frac{(c_1e^{2u+v+y}+c_2)+\al e^{u+y}+\be e^{u+v}\ + rs}
{(c_1e^{2u+v+y}+c_2)+\al e^{u+v}+\be e^{u+y}\ + rs}\right]=0.\quad 
\]
where the equation holds modulo \eqref{n-expCD} and its shifts. Moreover,

\[
 \bigl[D_{\tau_1},D_{\tau_2}\bigr]u=0\quad  mod\text{\eqref{n-expCD}}\quad   \text{and its shifts}.
\]

\end{theorem}

We now turn to the B\"acklund transformation~\eqref{B-expCD2}. Eliminating the derivatives in this case gives
\begin{equation}
\begin{aligned}
 &\left(\sqrt{c_1}+\sqrt{c_2}\,e^{-v-u}\right)
 \sqrt{e^{2v}+2 \al e^{v+u}+e^{2u}}-\left(\sqrt{c_1}+\sqrt{c_2}\,e^{-z-y}\right)
 \sqrt{e^{2z}+2 \al e^{z+y}+e^{2y}}\\
 &\quad -\left(\sqrt{c_1}+\sqrt{c_2}\,e^{-y-u}\right)
 \sqrt{e^{2y}+2 \be e^{y+u}+e^{2u}} +\left(\sqrt{c_1}+\sqrt{c_2}\,e^{-z-v}\right)
 \sqrt{e^{2z}+2 \be e^{z+v}+e^{2v}}=0.
\end{aligned}
\end{equation}

To compare the two polynomials, set \(c_1=c_2=1\) in the multi-quadratic quad equation \eqref{Q-expCD} obtained from \eqref{B-expCD1}. Inversion of the two opposite vertices then gives

\[
Q_{\mathrm{exCD2}}(U,V,Y,Z;\al,\be)=U^2Z^2Q_{\mathrm{exCD}}\!\left(\frac1U,V,Y,\frac1Z;\al,\be,\ c_1=c_2=1\right).
\]

where $Q_{\mathrm{exCD2}}$ is obtained from the Bianchi diagram for \eqref{B-expCD2}.

For the degenerate chain~\eqref{B-expCD3}, let $\al_1,\al_2$ be the parameters on the horizontal edges and $\be_1,\be_2$ their vertical counterparts. The corresponding edge relation is written as

\begin{equation}
\begin{aligned}
 &\al_1e^{(v+ u)/2}+\al_2e^{-(v+ u)/2} + \al_1e^{(z+ y)/2}+\al_2e^{-(z+ y)/2}\\
 &-\be_1e^{(y+ u)/2}\be_2e^{-(y+ u)/2} - \be_1e^{(z+ v)/2}-\be_2e^{-(z+ v)/2}=0.
\end{aligned}
\end{equation}

In the exponential variables, the stated factorization of the eliminant is
\[
P=(UZ-VY)^2\,Q_{\mathrm{deg}}\,G_{\mathrm{deg}},
\]
The component obtained by squaring is
\[
Q_{\mathrm{deg}}=\left[Z(\al_1^2V+\be_1^2Y)-U(\be_2^2V+\al_2^2Y)\right]^2-4VY(\al_1\be_1Z-\al_2\be_2U)^2.
\]
When the two coefficients in each direction coincide, this component agrees with the following
specialization of $Q_{\mathrm{exCD}}$.
\[
\left.Q_{\mathrm{deg}}\right|_{\substack{\al_2=\al_1\\\be_2=\be_1}}=\left.Q_{\mathrm{exCD}}\right|_{\substack{c_1=c_2=0\\\al=\al_1^2,\ \be=\be_1^2}}.
\]

\subsection{The Krichever--Novikov equation}\label{sec:quad-kn}

Consider the Krichever--Novikov (KN) equation
\begin{equation}\label{KN}
    u_t=u_{xxx}-\frac{3}{2}\,\frac{u^{2}_{xx}}{u_{x}}+\frac{q(u)}{u_{x}}
\end{equation}

Its B\"acklund transformation has a symmetric biquadratic right-hand side. The discriminant condition for the prescribed polynomial $q(u)$ in \eqref{KN} constrains the coefficients of the chain
\begin{equation}\label{B-KN}
v_xu_x=a_1v^2u^2+a_2vu(v+u)+a_3(v^2+u^2)+a_5vu+a_6(v+u)+a_9,
\end{equation}

Writing $q(\xi)=\sum_{j=0}^4 b_j\xi^j$, the coefficient conditions are
\begin{equation}\label{KN-B-coeff}
    \begin{aligned}
        b_4&=-6a_1a_3+3a_2^2/2,\\
        b_3&=-6a_1a_6-6a_2a_3+3a_2a_5,\\ 
        b_2&=-6a_1a_9-3a_2a_6-6a_3^2+3a_5^2/2,\\ 
        b_1&=-6a_2a_9-6a_3a_6+3a_5a_6,\\
        b_0&=-6a_3a_9+3a_6^2/2.
\end{aligned}
\end{equation}

\begin{prop}
The Krichever--Novikov equation~\eqref{KN} is consistent with the B\"acklund transformation~\eqref{B-KN}. Namely, the derivations satisfy $\left[D_t,D_{x}\right]=0.$
\end{prop}

Eliminating the derivatives from the four B\"acklund relations gives a polynomial with two multiaffine factors. The selected component has the form introduced in \eqref{eq:quad-kn},

\begin{equation}\label{Q-KN}\begin{aligned}
 Q_{KN}={}&c_1 u v y z+c_2 (u v y+u v z+u y z+v y z)+c_3 (u v+y z)\\ &+c_4 (u y+v z)+c_5 (u z+v y)+c_6 (u+v+y+z)+c_7=0.
\end{aligned}\end{equation}

The horizontal and vertical edge biquadratic polynomials are
\begin{align}\label{eq:kn-biquads}
 r(u,v)&=\partial_y Q_{KN}\,\partial_z Q_{KN}-Q_{KN}\,\partial_y\partial_z Q_{KN},\\
 s(u,y)&=-\partial_v Q_{KN}\,\partial_z Q_{KN}+Q_{KN}\,\partial_v\partial_z Q_{KN}.
\end{align}
Multiaffinity makes each right-hand side independent of the two vertices with respect to which the derivatives are taken. The B\"acklund transformation and discrete symmetry can be written in terms of the biquadratic polynomials as
\begin{equation}\label{t-KN}\begin{aligned}
 u_xv_x&=r(u,v),&u_xy_x&=s(u,y),\\
 u_{\tau_1}&=\frac{r(u,v)}{v-w}-\frac12\partial_v r(u,v),&
 u_{\tau_2}&=\frac{s(u,y)}{y-u_{0,-1}}-\frac12\partial_y s(u,y).
\end{aligned}\end{equation}
These expressions give the two lattice flows entering the compatibility statement.

By the explicit substitutions described in Section \ref{sec-flows} we obtain the following theorem
\begin{theorem}\label{thm:kn}
The Krichever--Novikov equation~\eqref{KN}, the B\"acklund transformation~\eqref{B-KN}, and its three-point symmetry~\eqref{t-KN} are pairwise consistent. The multiaffine relation~\eqref{Q-KN} is invariant under either lattice flow, and the two flows commute on its solutions. Namely,
\[
\frac{d}{d\tau}\left[ Q_{KN}\right]=0.\quad 
\]
This identity holds modulo \eqref{Q-KN} and its shifts. In addition,
\[
 \bigl[D_{\tau_1},D_{\tau_2}\bigr]u=0\quad  mod(\text{\eqref{Q-KN}})\quad   \text{and its shifts}.
\]

\end{theorem}

\subsection{Schwarzian KdV}\label{sec:quad-degkn}

The zero-quartic specialization of \eqref{KN} is the Schwarzian KdV equation
\begin{equation}\label{ShKdV}
    u_t=u_{xxx}-\frac{3u_{xx}^2}{2u_x}.
\end{equation}

The general B\"acklund transformation is the square of a bilinear polynomial. The edge relation becomes
\begin{equation}\label{BB-ShKdV}
   v_xu_x=(avu+bv+cu+d)^2.
\end{equation}

The nonautonomous transformation $\tilde u_n=\frac{2 au_n-A+b+c}{2au_n+A+b+c}(-\frac{A+b-c}{A-b+c})^n,\;A=\sqrt{(b+c)^2-4ad}$ gives the simpler B\"acklund transformation

\begin{equation}\label{B-ShKdV}
   v_xu_x=\alpha(u-v)^2.
\end{equation}

\begin{prop}
The Schwarzian KdV evolution~\eqref{ShKdV} is consistent with the B\"acklund transformation \eqref{B-ShKdV}. Namely,
$\left[D_t,D_{x}\right]=0.$
\end{prop}

For the elimination on the square, introduce the linear edge factors

\[
r(x,y)=\sqrt{\al}(x-y)\quad s(x,y)=\sqrt{\be}(x-y).
\]

The product of the two horizontal edge relations and the product of the two vertical ones both contain the same four derivatives. Equating the products eliminates them and gives
\begin{equation}
r(v,u)^2r(z,y)^2-s(y,u)^2s(z,v)^2=0.
\end{equation}

This difference of squares factors as $(r(v,u)r(z,y)-s(y,u)s(z,v))(r(v,u)r(z,y)+s(y,u)s(z,v))=0$. Each factor is a cross-ratio equation. With an appropriate choice of parameter sign and normalization, it gives $Q_1$ with zero parameter. Thus, we obtain the following equation:

\begin{equation}\label{Q-ShKdV}
    Q_{Sh}=\al (u-v)(y-z)-\be (u-y)(v-z).
\end{equation}

For this multiaffine quad equation, define the edge biquadratics $r,s$ by
\begin{align}\label{eq:sh-biquads}
 (\be-\al)r(u,v)&=\partial_y Q_{Sh}\,\partial_z Q_{Sh}-Q_{Sh}\,\partial_y\partial_z Q_{Sh},\\
 (\al-\be)s(u,y)&=\partial_v Q_{Sh}\,\partial_z Q_{Sh}-Q_{Sh}\,\partial_v\partial_z Q_{Sh}.
\end{align}

The horizontal edge biquadratic is proportional to $(u-v)^2$. Substitution into the three-point formula gives
\begin{equation}\label{t-ShKdV}
    u_{\tau_1}=\frac{r(u,v)}{v-w}-\frac12\partial_v r(u,v)=\alpha \frac{(u-w)(u-v)}{v-w}
\end{equation}
 This gives the the three-point symmetry for Schwarzian KdV \eqref{V-ShKdV} with linear term absent.

 By the explicit substitutions described in Section \ref{sec-flows} we obtain the following theorem
\begin{theorem}\label{thm:schwarzian}
The Schwarzian KdV equation~\eqref{ShKdV}, its B\"acklund transformation~\eqref{B-ShKdV}, and the three-point symmetry~\eqref{t-ShKdV} are pairwise consistent. On solutions of the multiaffine equation~\eqref{Q-ShKdV}, the horizontal and vertical flows preserve the quad relation and commute. In particular,

\[
\frac{d}{d\tau}\left[ Q_{Sh}\right]=0.\quad 
\]

The equality holds modulo \eqref{Q-ShKdV} and its shifts, as does the commutator identity

\[
 \bigl[D_{\tau_1},D_{\tau_2}\bigr]u=0\quad  mod\text{\eqref{Q-ShKdV}}\quad   \text{and its shifts}.
\]

\end{theorem}

\subsection{The potential Gardner equation}\label{sec:quad-mmkdv}

We call the following equation the potential Gardner equation:
\begin{equation}\label{pGar}
    u_t=u_{xxx}-\frac{3u_{xx}^2}{4u_x}+a_1u_x^{3/2}-3a_2^2u_x^2
\end{equation}

The transformation $v^2=u_x$ converts it into the Gardner equation. The potential Gardner equation admits the following B\"acklund transformation when $a_1=0$:

\begin{equation}\label{B-pGar}
    \sqrt{v_x}+\sqrt{u_x}=a_2\sqrt{(v-u)^2-\al}
\end{equation}

For general $a_1$, the B\"acklund transformation is expressed in terms of a solution of an ODE. We do not consider this case here.

\begin{prop}
The potential Gardner equation~\eqref{pGar} with $a_1=0$ is consistent with the B\"acklund transformation~\eqref{B-pGar}. Namely, $\left[D_t,D_{x}\right]=0.$
\end{prop}

Eliminating the derivatives on a Bianchi square gives
\begin{equation}
 \sqrt{(u-v)^2-\al}
 +\sqrt{(y-z)^2-\al}-\sqrt{(u-y)^2-\be}-\sqrt{(v-z)^2-\be}=0.
\end{equation}

Denote the four radicands by
\[
A=(v-u)^2-\al,\quad B=(z-y)^2-\al,\quad
C=(y-u)^2-\be,\quad D=(z-v)^2-\be.
\]
Successive squaring of $\sqrt A+\sqrt B=\sqrt C+\sqrt D$ yields
\begin{equation}
P=\bigl[(A+B-C-D)^2-4AB-4CD\bigr]^2-64ABCD=0.
\end{equation}
To express the factorization, introduce three linear combinations of the vertex values,
\[
X=u-v+y-z,\qquad Y=u+v-y-z,\qquad T=u-v-y+z.
\]
The exact factorization is
\begin{equation}P=4T^2Q_G,\end{equation}
and the nontrivial multi-quadratic factor is written as

\begin{equation}\label{Q-pGar}
Q_G=(\be X^2-\al Y^2)\bigl[X^2-Y^2-4(\al-\be)\bigr]+(\al-\be)T^2(X^2-Y^2).
\end{equation}

The horizontal radicand determines the three-point characteristic
\begin{equation}\label{t-pGar}
    u_\tau=\frac{R(v,u,w)-r(u,v)r(u,w)}{v-w}\quad R(v,u,w)=r(u,v)^2-\frac{1}{2}\partial _v(r(u,v)^2)(v-w)
\end{equation}
Here $r=(p^2-\al)^\frac12$ and it's general form of discrete symmetry \eqref{V-pGar} with $k_2=0$. To solve for the opposite vertex, introduce the initial edge differences
\[
p=v-u,\quad q=y-u,
\]
Solving the multi-quadratic equation $Q_G=0$ for $z$ and selecting one branch gives
\begin{equation}\label{n-pGar}
z=u+\frac{(\al-\be)(p-q)}{\al+\be-2pq-2\sqrt{(p^2-\al)(q^2-\be)}}.
\end{equation}

Here $r=(p^2-\al)^\frac12$ and $s=(q^2-\be)^\frac12$. The outgoing horizontal and vertical roots satisfy $r_z=((z-y)^2-\al)^\frac12$ and $s_z=((z-v)^2-\be)^\frac12$. Substituting the vertex update turns both radicands into perfect squares. The resulting rational map is

\begin{equation}\begin{aligned}
\Delta&=\al+\be-2pq-2rs,\\
z &=u+\frac{(\al-\be)(p-q)}\Delta,\\
r_z&=s+\frac{(\al-\be)(r+s)}{\Delta},\\
 s_z&=r-\frac{(\al-\be)(r+s)}{\Delta}.
\end{aligned}\label{map-pGar}\end{equation}

By the explicit substitutions described in Section \ref{sec-flows} we obtain the following theorem
\begin{theorem}\label{thm:gardner}
The potential Gardner equation~\eqref{pGar} in case $a_1=0$, the B\"acklund transformation~\eqref{B-pGar}, and the three-point symmetry~\eqref{t-pGar} are pairwise consistent. The two lattice flows preserve the nonlinear superposition formula~\eqref{n-pGar} and commute by virtue of \eqref{map-pGar}. Namely,

\[
\frac{d}{d\tau}\left[z-u-\frac{(\al-\be)(p-q)}{\al+\be-2pq-2\sqrt{(p^2-\al)(q^2-\be)}}\right]=0.\quad mod\text{\eqref{map-pGar}}\quad   \text{and its shifts}.
\]

\[
 \bigl[D_{\tau_1},D_{\tau_2}\bigr]u=0\quad  mod\text{\eqref{map-pGar}}\quad   \text{and its shifts}.
\]

\end{theorem}

\subsection{The elliptic Calogero--Degasperis equation}\label{sec:quad-rat2}

Consider the elliptic Calogero--Degasperis equation
\begin{equation}\label{elCD}
    u_{t}= u_{xxx}- \frac{3}{8}\frac{\big((q(u)+u_x^2)_{x}\big)^2}{u_x\,(q(u)+u_x^2)}+\frac{1}{2} q''(u)\,u_x
\end{equation}

Its B\"acklund transformation has the following form:
\begin{equation}\label{B-elCD}
    \ArcSinh\frac{v_x}{\sqrt{q(v)}}-c\ArcSinh\frac{u_x}{\sqrt{q(u)}}=\operatorname{arccosh}\frac{cA(v,u)}{\sqrt{q(v)}\sqrt{q(u)}},\ c=\pm1.
\end{equation}

We consider five  forms of $q(u)$. For each form, we list the functions $A(u,v)$ that define a B\"acklund transformation for \eqref{elCD}.

For $q(u)=u(u-1)(u-k)$, there are four admissible edge functions $A$.
\begin{align}
A_1(u,v)&=\al(u-v)^2-uv/2(u+v-2k-2)-k/2(u+v)\\
A_2(u,v)&=\al(k(u+v-1)-uv)^2-k/2(u^2+v^2-u-v)+uv/2(u+v-2),\\
A_3(u,v)&=\al(uv-k)^2+uv/2(u+v-2)+k/2(u+v-2uv),\\
A_4(u,v)&=\al(uv-u-v+k)^2+u^2/2(v-1)+v^2/2(u-1)+k/2(u+v-2uv)
\end{align}

For $q(u)=u^2-1$, the two possible choices of $A$ are

\begin{equation}
    A_5(u,v)=(u-v)^2\al-uv+1\quad A_6(u,v)=(u+v)^2\al -uv-1
\end{equation}

For $q(u)=u^2$, the two functions $A$ are

\begin{equation}
    A_7(u,v)=\al(uv+1)^2+uv\quad A_8(u,v)=\al(v+u)^2-uv
\end{equation}

For $q(u)=u$, there is a single choice of $A$,

\begin{equation}
    A_9(u,v)=\al(u-v)^2/2-u/2-v/2
\end{equation}

Finally, for $q(u)=1$, the two functions $A$ are
\begin{equation}
    A_{10}(u,v)=\al(u+v)^2+1,\quad A_{11}(u,v)=\al(u-v)^2-1.
\end{equation}
\begin{remark}
    We omit several cases from the article \cite{Garif_2023}, as they are Darboux integrable and therefore possess pseudoconstants. In accordance with our convention, such cases are excluded from consideration.
\end{remark}
\begin{prop}
For each of the functions $A_i(u,v)$ listed above, the elliptic Calogero--Degasperis equation~\eqref{elCD} is consistent with the B\"acklund transformation~\eqref{B-elCD} with the corresponding polynomials $q$ and $A(u,v)$. Namely, $\left[D_t,D_{x}\right]=0.$ Here $D_{x}$ is defined by the corresponding $A(u,v)$.
\end{prop}

We give the compatibility calculation for the third case, with $A=A_3$ and $q(u)=u(u-1)(u-k)$. The horizontal edge root is defined by the identity

\[
4\left(A(u,v)^2-q(u)q(v)\right)
=\Delta(u,v)^2r(u,v)^2,\quad \Delta(x,y)=xy-k.
\]

Here $r$ is a signed square root on the horizontal edge, and $s$ denotes its vertical counterpart with parameter $\beta$. Combining the four B\"acklund relations to eliminate the transformed derivatives and applying $\cosh$ gives

\[
\begin{aligned}
&4\left[A_\beta(u,y)A_\beta(v,z)-A_\alpha(u,v)A_\alpha(y,z)\right]+\\
&+\Delta(u,y)\Delta(v,z)\sqrt{s(u,y)s(v,z)}-\Delta(u,v)\Delta(y,z)\sqrt{r(u,v)r(y,z)}=0.
\end{aligned}
\]

Eliminating the radicals now gives a polynomial with a multi-quadratic factor. To write that factor, put
\[
\delta=\alpha-\beta,\qquad \Sigma=(u-z)(v-y),
\]
and define
\[
C(a,b;c,d)=ab(c+d-k-1)-cd(a+b-k-1)+k(a+b-c-d).
\]
The multi-quadratic quad polynomial is
\begin{equation}\label{Q-elCD}
    \begin{aligned}
Q_{\mathrm{elCD}}\coloneqq{}&\delta\Bigl[2\alpha(uv-yz)C(u,v;y,z)-2\beta(uy-vz)C(u,y;v,z)\\
&\qquad+\Sigma\bigl(4\alpha\beta(uy-k)(vz-k)-(u-y)(v-z)\bigr)\Bigr]-\beta(2\alpha+1)(2\alpha k+1)\Sigma^2.
\end{aligned}
\end{equation}
Below, let $Q$ denote the polynomial $Q_{\mathrm{elCD}}$ in \eqref{Q-elCD}. Its discriminant in $y$ factors into the biquadratics on the two edges adjacent to $v$, namely,
\[
Q_y^2-2QQ_{yy}=\kappa_i^2(z-u)^2r(u,v)^2s(v,z)^2,
\]
where $\kappa_i=\delta$ for case 3. The horizontal edge root from this factorization determines the discrete symmetry
\begin{equation}\label{t-elCD}
    u_\tau=\frac{R(v,u,w)-r(u,v)r(u,w)}{v-w}\quad R(v,u,w)=r(u,v)^2-\frac{1}{2}\partial _v(r(u,v)^2)(v-w)
\end{equation}

One could solve the multi-quadratic relation for the fourth vertex and substitute the result as in the preceding cases. Here, it is convenient to eliminate the neighbouring vertices using identities among the derivatives of the quad polynomial. The computation has two parts. We first reduce the horizontal derivative to an expression on the central square. We then verify the equation by explicit computation.

Write $Q_i$ for the selected polynomial in case $i$. Hold its first and third vertex arguments fixed and introduce the transformed polynomial
\[
\Phi_i(U,W)=U^2W^2 Q_i\left(u,y+\frac1U,y,u-\frac1W\right)
\]

For the displayed case, the resulting $\Phi_i$ has total degree two in $U,W$. The following equation holds identically for every polynomial of total  degree at most  two:

\[
\begin{aligned}
&\bigl[\Phi_U(U,W)+\Phi_U(U_0,W_0)\bigr](U-U_0)\\
&\quad+\bigl[\Phi_W(U,W)+\Phi_W(U_0,W_0)\bigr](W-W_0)\\
&=2\bigl[\Phi(U,W)-\Phi(U_0,W_0)\bigr].
\end{aligned}
\]

We also have the following gradient relation and its shifted versions:

\[
Q_z^2-2QQ_{zz}=\kappa_i^2(y-v)^2r(u,v)^2s(u,y)^2,
\]

\[
Q_v^2-2QQ_{vv}=\kappa_i^2(z-u)^2r(y,z)^2s(u,y)^2,
\]

On $Q=0$, the discriminants give squared identities for $Q_v$ and $Q_z$. After choosing signs for the edge roots, the first-derivative relations are written as

\begin{equation}
    Q_v=\kappa_i(z-u)r(y,z)s(u,y)\quad Q_z=\kappa_i(y-v)r(u,v)s(u,y)
\end{equation}

On the quad relation, differentiating $\Phi$ gives the following relations involving the derivatives of the vertex polynomial:

\[
Q_v=-(u-z)^2\Phi_U,\qquad
Q_z=(v-y)^2\Phi_W.
\]

Combining the equations above and performing some algebraic manipulations, we obtain

\[
(u-z)\zeta-(v-y)\eta=r+r_z.
\]

where 

\[
\eta=\frac{r_u+r}{w-v},
\qquad
\zeta=\frac{r_y+r_z}{j-z}.
\]

The polynomials in the three-point discrete symmetry have the following form:

\[
R(v,u,w)=r^2-\frac{1}{2}\,\partial_v(r^2)(v-w),
\]
\[
R(j,y,z)=r_z^2-\frac{1}{2}\,\partial_z(r_z^2)(z-j).
\]

Substituting these expressions into the horizontal characteristic gives

\[
Q_u u_\tau+Q_y y_\tau={}-\frac12\left[Q_u\partial_v(r^2)+Q_y\partial_z(r_z^2)\right]-Q_u r\eta-Q_y r_z\zeta.
\]

The signed edge gradients on the central face also satisfy

\[
Q_u r=-\frac{v-y}{u-z}Q_y r_z.
\]

The last identity relates the coefficients of $\eta$ and $\zeta$. Applying it allows us to eliminate the variables at the two left vertices. Modulo the left and central face equations, this gives
\[
Q_u u_\tau+Q_y y_\tau\equiv-\frac12\left[Q_u\partial_v(r^2)+Q_y\partial_z(r_z^2)\right]+\frac{Q_u r^2}{v-y}-\frac{Q_y r_z^2}{u-z}.
\]

The right-hand side now involves only the central vertices. The same calculation on the rightneighbouring square gives the contribution from $v$ and $z$ in the form

\[
Q_v v_\tau+Q_z z_\tau\equiv\frac12\left[Q_v\,\partial_u(r^2)+Q_z\,\partial_y(r_z^2)\right]+\frac{Q_z r_z^2}{v-y}-\frac{Q_v r^2}{u-z}.
\]

Adding the two contributions expresses $Q_\tau$ entirely in terms of the central face.

\begin{equation}
    Q_\tau=\frac12\left[Q_v\,\partial_u(r^2)+Q_z\,\partial_y(r_z^2)\right]+\frac{Q_z r_z^2}{v-y}-\frac{Q_v r^2}{u-z}-\frac12\left[Q_u\partial_v(r^2)+Q_y\partial_z(r_z^2)\right]+\frac{Q_u r^2}{v-y}-\frac{Q_y r_z^2}{u-z}
\end{equation}

After the denominators are cleared, the numerator is divisible by the central quad polynomial. Consequently, the $\tau$ flow preserves the central relation, as expressed by

\begin{equation}
    D_\tau Q=K Q=0 \quad mod(Q)
\end{equation}

where $K$ denotes the factor multiplying $Q$. This proves the consistency of the discrete symmetry with the multi-quadratic quad equation.

For commutativity of the $\tau_1,\tau_2$ flows, take $y$ as the central vertex of a three-by-three array. Denote its four face equations by

\[
\begin{aligned}
Q&=Q(u,v,y,z)=0,&\qquad P&=P(w,u,j,y)=0,\\
\widehat Q&=Q(y,z,\hat y,\hat z)=0,&\qquad \widehat P&=P(j,y,\hat j,\hat y)=0.
\end{aligned}
\]

We retain the five vertices $u,j,y,z,\hat y$ and the four edge roots adjacent to $y$. For each corner, the corresponding face polynomial is quadratic in the corner variable, so its first derivative is affine in that variable. Solving the signed discriminant identities for the four corners gives

\[
\begin{aligned}
w&=\frac{-\delta(u-j)r_y s-\left.P_w\right|_{w=0}}{P_{ww}},&
v&=\frac{-\delta(u-z)r_z s-\left.Q_v\right|_{v=0}}{Q_{vv}},\\[4pt]
\hat j&=\frac{\delta(j-\hat y)r_y s_{\hat y}-\left.\widehat P_{\hat j}\right|_{\hat j=0}}{\widehat P_{\hat j\hat j}},
&\hat z&=\frac{\delta(z-\hat y)r_z s_{\hat y} -\left.\widehat Q_{\hat z}\right|_{\hat z=0}}{\widehat Q_{\hat z\hat z}}.
\end{aligned}
\]

The same identities determine the remaining edge roots. For example, on the two lower facesone obtains

\[
\begin{aligned}
r_u&=\frac{P_j}{\delta(w-y)s},&
s_j&=-\frac{P_u}{\delta(w-y)r_y},\\[3pt]
r&=\frac{Q_z}{\delta(v-y)s},&
s_z&=-\frac{Q_u}{\delta(v-y)r_z}.
\end{aligned}
\]

After these substitutions, the expression $\left[D_{\tau_1},D_{\tau_2}\right]y$ depends only on the five retained vertices and the edge roots $r_y,r_z,s,s_{\hat y}$. Each pair consisting of a neighbouring vertex and its incident root lies on a conic.

To carry out the remaining cancellation, parametrize each conic by a line through a known point. For example, consider the conic
\begin{center}
$
r_z^2=L^2z^2+2Bz+C^2.
$
\end{center}
It contains the point $(0,C)$. The slope of the line joining that point to $(z,r_z)$ is
\begin{center}
$
p_+=\frac{r_z-C}{z}.
$
\end{center}
Substituting $r_z=C+p_+z$ into the conic and cancelling the common factor $z$ gives the vertex and its root in the form
\begin{center}
$
z=\frac{2(B-C p_+)}{p_+^2-L^2},
\qquad
r_z=C+p_+z.
$
\end{center}

Use the analogous parametrization for the other three edges. The reduced commutator is then rational in the four slope parameters and the central data. Explicit calculation then yields the following equation by virtue of the multi-quadratic quad equation and its shifts:

\begin{equation}
    \left[D_{\tau_1},D_{\tau_2}\right]y=0
\end{equation}

It remains to relate this calculation to the other choices of $A_i$. For the cubic $q$, cases $1,2,3,4$ are related by changes of the B\"acklund parameters. Up to a normalization depending only on those parameters, the quad polynomials satisfy

\[
Q_i(u,v,y,z;\alpha,\beta)=\gamma_i(\alpha)^2\gamma_i(\beta)^2Q_{elCD}\bigl(u,v,y,z;\mu_i(\alpha),\mu_i(\beta)\bigr).
\]

For case 1, the map is $\mu(x)=-\frac{1}{4x}$. The maps for cases 2 and 4 are $\frac{x}{1-2kx}$ and $\frac{x}{1-2x}$, respectively. Case 3 is the reference polynomial, with the identity parameter map.

For $q(x)=x(x-1)(x-k)$, it is therefore sufficient to carry out the calculation for one representative. For the other choices of $q(x)$, the transformations depend on the family. In cases 5 and 6, the parameter change is $\mu(x)=\frac{x}{1-2x}$. For the pair described as cases 7 and 8, the  relation is $Q_{7}(u,v,y,z;\alpha,\beta)=v^2y^2\,Q_{8}\!\left(u,\frac1v,\frac1y,z;-\alpha,-\beta\right).$ Such transformations allow us to perform the explicit computation for only one case for each form of $q(u)$. One can carry out described above calculations and obtain the following theorem.

\begin{theorem}\label{thm:elliptic-cd}
For each edge function $A_i$, the elliptic CD equation~\eqref{elCD}, the corresponding dressing chain~\eqref{B-elCD}, and its three-point symmetry~\eqref{t-elCD} are pairwise consistent. Each discrete symmetry flow preserves the associated multi-quadratic quad equation, and the horizontal and vertical flows commute modulo that relation and its shifts. Write $Q_{elCD,i}$ for the selected polynomial in case $i$, with \eqref{Q-elCD} as the representative for case 3. Namely,
\[
\frac{d}{d\tau}\left[Q_{elCD,i}\right]=0.\quad mod(Q_{elCD,i})\quad   \text{and its shifts}.
\]

\[
 \bigl[D_{\tau_1},D_{\tau_2}\bigr]u=0\quad  mod(Q_{elCD,i})\quad   \text{and its shifts}.
\]

\end{theorem}

\section{Integrability of the multi-quadratic quad equations}\label{sec:integrability}

The theorems of Section~\ref{sec:quads} establish the compatibility of the quad equations with the discrete symmetries. They do not by themselves show that the multi-quadratic equations are multidimensionally consistent. In this section, we identify the multi-quadratic equations with known integrable equations and collect the resulting integrability properties. Most of the statements below are known in other notation; we indicate the sources.

\subsection{Identification with the Atkinson--Nieszporski list}

Atkinson and Nieszporski \cite{Atkinson} obtained multi-quadratic counterparts $Q4^*,\dots,H2^*$ of the ABS equations. These equations are 3D consistent. All of them except $Q4^*$ are connected to ABS equations by transformations of Bäcklund or Miura type
\cite[Table~2]{Atkinson}. Comparing the polynomials directly, we find the following.

\begin{prop}\label{prop:ident}
Write the quad equations of Section~\ref{sec:quads} in the notation $(u,v,y,z)=(u,\tilde u,\hat
u,\hat{\tilde u})$ of \cite{Atkinson}. Then:
\begin{enumerate}
\item $Q_{KdV}$ coincides with $H2^*$ with $p=\al$, $q=\be$;
\item $Q_{mKdV}$ coincides with $A1^*$ with $p=-\al$, $q=-\be$;
\item $Q_G$ coincides with $Q1^*$ with $p=\al$, $q=\be$;
\item $Q_{exCD}$ with $c_1c_2\neq0$ coincides with $A2^*$ after the scaling $U\mapsto (c_2/c_1)^{1/4}U$ of all vertex
variables, with $p=-\al/\sqrt{c_1c_2}$, $q=-\be/\sqrt{c_1c_2}$.
\end{enumerate}
In each case, the two polynomials have the same zero set.
\end{prop}

The equation $H2^*$ was known earlier. Adler and Veselov obtained it as the superposition principle for Bäcklund transformations of the KdV equation \cite{Adler_Veselov}, and it also appears in \cite{KN_2012} in connection with Yang--Baxter maps. Atkinson obtained $A2^*$ as the superposition principle for Bäcklund transformations of Hirota's discrete KdV equation \cite{Atkinson_2012}. $A1^*$ appeared implicitly in earlier work; see \cite{Atkinson}. Thus, the quad equations obtained in Section~\ref{sec:quads} for KdV, mKdV, potential Gardner and exponential Calogero--Degasperis are known equations in a new context. Now they arise from the Bianchi diagrams of the corresponding KdV-type equations.

By Proposition~\ref{prop:ident} and \cite{Atkinson}, these four equations are 3D consistent. They are also related to ABS equations: $H2^*$ to H1, $A1^*$ and $Q1^*$ to $H3_{\delta=0}$, and $A2^*$ to A2 \cite[Table~2]{Atkinson}. For the reader's convenience, we write the first two relations in the present notation. Up to point transformations of the potential and a reparametrisation of the lattice parameters, the first is the H1 entry of \cite[Table~2]{Atkinson}, which goes back to \cite{Adler_Veselov}, and the second is the $H3_{\delta=0}$ entry for $A1^*$ and $Q1^*$ there.

\begin{prop}\label{prop:miura-kdv}
Let $\phi$ satisfy H1, $(\phi_{12}-\phi)(\phi_1-\phi_2)=\al-\be$. Then the system
\begin{equation}\label{miura-kdv}
u+T_iu=(T_i\phi-\phi)^2-\al_i,\qquad \al_1=\al,\ \al_2=\be,
\end{equation}
is compatible, and its solutions satisfy $Q_{KdV}=0$. The edge roots of Section~\ref{sec:quad-kdv} are $r=\phi_1-\phi$ and $s=\phi-\phi_2$.
\end{prop}

\begin{proof}
System~\eqref{miura-kdv} is compatible if and only if the alternating sum of its right-hand sides around a square vanishes. By the computation in Section~\ref{sec:preliminaries}, this sum equals $2(\phi_{12}-\phi)(\phi_1-\phi_2)-2(\al-\be)$. Substituting $r=\phi_1-\phi$ and $s=\phi-\phi_2$ into~\eqref{map-KdV} and using $\phi_{12}-\phi=(\al-\be)/(r+s)$ gives $r_z=\phi_{12}-\phi_2$ and $s_z=\phi_1-\phi_{12}$. The vertex formula in~\eqref{map-KdV} then follows from~\eqref{miura-kdv}.
\end{proof}

\begin{prop}\label{prop:miura-mkdv}
Let $A^2=\al$, $B^2=\be$, and let $\psi$ satisfy the lattice mKdV equation $B(\psi\psi_1-\psi_2\psi_{12})=A(\psi\psi_2-\psi_1\psi_{12})$. This equation becomes $H3_{\delta=0}$ under $\psi\mapsto i^{n+m}\psi$. Then the systems
\[
u+T_iu=\frac{A_i}{2}\left(\frac{T_i\psi}{\psi}-\frac{\psi}{T_i\psi}\right),\qquad
T_iu-u=\frac{A_i}{2}\left(\frac{T_i\psi}{\psi}+\frac{\psi}{T_i\psi}\right),\qquad A_1=A,\ A_2=B,
\]
are compatible. Their solutions satisfy $Q_{mKdV}=0$ and $Q_G=0$, respectively. With $T_i\psi/\psi=e^{\theta_i}$, the radicals in~\eqref{B-mKdV} and~\eqref{B-pGar} become $A_i\cosh\theta_i$ and $A_i\sinh\theta_i$, respectively.
\end{prop}

These relations are lattice analogues of the Miura-type relations $u\sim\phi_x$ for KdV and $u\sim(\log\psi)_x$ for mKdV.

\subsection{Edge variables and Yang--Baxter maps}

The rational maps~\eqref{map-KdV}, \eqref{map-mKdV} and~\eqref{map-pGar} are instances of the single-valued vertex--edge reformulation of multi-quadratic equations in \cite[Section~5]{Atkinson}. The relation between multi-quadratic equations and Yang--Baxter maps is also known. Equations of this class arise as potentials of the quadrirational Yang--Baxter maps \cite{ABS_YB,PSTV} on the edges \cite{KN_2012,KN_2011,KN_2018}, \cite[Section~7.4]{Atkinson}. In the present examples, the new edge roots do not depend on the vertex variables once suitable edge coordinates are chosen. Hence the maps induce Yang--Baxter maps $(X,Y)\mapsto(X',Y')$ as follows.
\begin{itemize}
\item KdV: with $X=r$, $Y=s$, the map is
$X'=Y+\frac{\al-\be}{X+Y}$, $Y'=X-\frac{\al-\be}{X+Y}$. This is the Adler map
\cite{Adler_1993,PTV}, i.e.\ $H_V$ in the notation of \cite{PSTV}.
\item mKdV: put $X=(u+v)+r$ and  $Y=(u+y)-s$.
Equation then gives the map
\[
X'=Y\,\frac{XY+\al}{XY+\be},\qquad Y'=X\,\frac{XY+\be}{XY+\al}.
\]
After the rescaling $X=\al x$, $Y=\be y$, this is the map $H_{III}^B$ of \cite{PSTV} with parameters
$(p,q)=(\al,\be)$.
\item Exponential Calogero--Degasperis \eqref{B-expCD1} with $c_1=C^2$: put $X=r+Ce^{u+v}$ and
$Y=s+Ce^{u+y}$. Then
\[
X'=\frac{Y\bigl(XY+\tfrac{\be}{C}X+\tfrac{\al}{C}Y+c_2\bigr)+\tfrac{c_2(\be-\al)}{C}}{XY+\tfrac{\be}{C}(X+Y)+c_2},\qquad
Y'=\frac{X\bigl(XY+\tfrac{\al}{C}Y+\tfrac{\be}{C}X+c_2\bigr)+\tfrac{c_2(\al-\be)}{C}}{XY+\tfrac{\al}{C}(X+Y)+c_2}.
\]
This map is involutive and satisfies the Yang--Baxter equation.
\end{itemize}

\subsection{The elliptic Calogero--Degasperis case}
 For the elliptic Calogero--Degasperis equation the following identification holds.

\begin{prop}\label{prop:elcd-q4star}
Let $c\in\mathbb{C}\setminus\{0,\pm1,\pm i\}$ and put
\[
\sqrt{k}=\frac{1+c^2}{1-c^2},\qquad
w=\sqrt{k}\,\frac{x-c}{x+c},\qquad
2\sqrt{k}\,\al=\frac{c-p}{c+p},\qquad 
2\sqrt{k}\,\be=\frac{c-q}{c+q}.
\]
Then $x,\tilde x,\hat x,\hat{\tilde x}$ satisfy $Q4^*$ with parameters $p,q$ and constant $c$ in the notation of \cite{Atkinson} if and only if the corresponding $w,\tilde w,\hat w,\hat{\tilde w}$ satisfy $Q_{\mathrm{elCD}}=0$ with  $k$ and parameters $\al,\be$.
\end{prop}

The Möbius map in Proposition~\ref{prop:elcd-q4star} sends the branch points $\pm c,\pm c^{-1}$ of the $q4^*$ biquadratics to the branch points $0,1,k,\infty$ of the elliptic curve $y^2=u(u-1)(u-k)$. Cases 1, 2 and 4 are related to case 3 by the parameter changes described above, so all elliptic Calogero--Degasperis quad equations with cubic $q$ are equivalent to $Q4^*$.

Proposition~\ref{prop:elcd-q4star} also explains the earlier  observation that $Q_{\mathrm{elCD}}$ is 3D consistent. It also places the elliptic Calogero--Degasperis equation at the top of the multi-quadratic hierarchy, just as the Krichever--Novikov equation sits at the top of the multiaffine one through Q4 \cite{Adler_1998}. No transformation of Bäcklund or Miura type between $Q4^*$ and Q4 is known. \cite{Atkinson_2024}.

\section{Conclusion}\label{sec:conclusion}

We have studied quad equations associated with the KdV-type equations of the Meshkov--Sokolov list. For each equation that admits a Bäcklund transformation of the required form, we obtained the general Bäcklund transformation, derived the quad equation from the Bianchi permutability diagram, and found the general three-point discrete symmetry. For each equation, we proved that the KdV-type equation, its Bäcklund transformation and the discrete symmetry are pairwise compatible. We also proved that the discrete symmetries preserve the quad equations and that the flows in the two lattice directions commute, modulo the quad equations and their shifts.

The multiaffine quad equations obtained in this way are H1 for potential KdV, $Q1_{\delta=0}$ for Schwarzian KdV and Q4 for the Krichever--Novikov equation. For the remaining equations, the Bianchi diagram yields multi-quadratic equations. For KdV, mKdV, potential Gardner and exponential Calogero--Degasperis, these are the equations $H2^*$, $A1^*$, $Q1^*$ and $A2^*$ of Atkinson and Nieszporski \cite{Atkinson}. Hence they are 3D consistent and related to ABS equations by transformations of Miura type. Thus, the integrable multi-quadratic quad equations arise naturally as nonlinear superposition principles for Bäcklund transformations of KdV-type equations, in the same way as the ABS equations do for their potential forms. The three-point symmetries obtained here appear to be the first systematic description of generalized symmetries for this class of equations. We also gave a simple method for reading off a three-point symmetry from the factorised vertex discriminant of a quad polynomial. This method works uniformly in the multiaffine and multi-quadratic cases.

For the elliptic Calogero--Degasperis equation with cubic $q$, the quad equation is the primary multi-quadratic model $Q4^*$, up to a Möbius change of variables. Thus the elliptic Calogero--Degasperis equation plays for the multi-quadratic equations the role that the Krichever--Novikov equation plays for the ABS list. 

The results can be extended in the following ways 
\begin{itemize}
    \item Is to find a direct transformation between $Q4^*$ and $Q4$ or prove its nonexistence.
    \item Find equations that  arise from reductions with respect to the negative symmetries associated with the obtained quad equations. Obtain such reductions  directly at the level of differential-difference chains. In some cases, eliminating derivatives from the Bianchi permutability diagram leads to polynomials of degree six in each variable.  And investigate whether these polynomials are 3D-consistent.
   \item Another open problem is to find the associated quad equations for cases in which the Bäcklund transformations cannot be expressed explicitly in terms of elementary functions. Further directions include proving the compatibility of the quad equations and nonlinear superposition principles with the $x$ dynamics, as well as constructing an analogous classification for the known lists of fifth-order equations.
   \item Of further interest are the construction of solutions to the obtained quad equations, the investigation of their continuous limits, and the study of the corresponding continuous limits of their solutions. We plan to address this in future work.
\end{itemize}

\section*{Acknowledgements}

This work was supported by the Foundation for the Advancement of Theoretical Physics and
Mathematics ``BASIS''.

I would like to thank Dr. V.Adler for  useful discussions.

\appendix
\section*{Appendix: determining-equation calculations}
In this appendix, we present intermediate calculations for solving the consistency condition for a KdV-type equation and a B\"acklund transformation. Each subsection derives the general form of the B\"acklund transformation for the corresponding equation. The consistency expression $D_t(V)$, where $V$ is as in \eqref{Backl-form}, depends on independent variables. Below, when we refer to the coefficient of $\boxed{u^n_{xxx}u^m_{xx}u^k_{x} }$, where any of $k,n,m$ may be zero, we mean that $D_t(V)$ is polynomial in the relevant variables and that this coefficient must vanish. Thus, solving the condition involving $D_t(V)$ reduces to a sequence of equations. Each $\bullet$ indicates that all preceding conditions have been solved and that we now impose the condition on $D_t(V)$ to refine $V$. Throughout the calculations below, we omit nonzero coefficients. 

\begin{remark}
    We assume the absence of pseudoconstants, i.e., that the only solutions of the  equations $D_x(c)=0 \quad (T-1)c=0$   are constants.
\end{remark}

\begin{remark}
    We omit the elliptic Calogero--Degasperis equation case due to the length of the calculations involved.
\end{remark}

\section{B\"acklund transformations }
\subsection{Potential KdV}

\begin{equation}
    u_t=u_{xxx}+u_x^2
\end{equation}

The PotKdV equation belongs to case A in \eqref{Appcond}, so its B\"acklund transformation has the form

\begin{equation}
    v_x=c(x,u,v)u_x+h(x,u,v)
\end{equation}

\begin{itemize}
 \item The coefficient of $\boxed{u_{xxx}}$ is $D\;c(x,u,v)=0$, so, by our convention, $c$ is constant.
    \item The coefficient of $\boxed{u_{xx}u_x}$ gives $-2c+2c^2+3(\partial_u+c\partial_v)^2h\Rightarrow$ $h=\frac{1-c}{12c}(c u+v)^2+v h_2(x,v-c u)+h_3(x,v-c u)$.
    \item The coefficient of $\boxed{u_{xx}u}$ after the substitution $v=w+cu$ is $-\frac{1}{3} (c-1) c \left(3 c \partial_wh_2(x,w)+c+1\right)$. There are two factors that can vanish.
    \begin{itemize}
        \item[$\triangledown$] If $c=1$, the coefficient of $\boxed{u_{xx}u}$ is $h_2(x,w) \left(3 \partial_wh_2(x,w)+2\right)$. Thus, this case is included in the case where the second factor vanishes. We therefore take $\left(3 c \partial_wh_2(x,w)+c+1\right)=0\Rightarrow h_2(x,w)=k(x)-\frac{(c+1) w}{3 c}$.
    \end{itemize}
    \item The coefficient of $\boxed{u_{xx}}$ gives $3ck'(x)\Rightarrow k$ is independent of x.
    \item The coefficient of $\boxed{u_x^2u}$ after the substitution $v=w+cu$ is $\frac{2c}{3}(c^2-1)$.
    \begin{itemize}
        \item[$\triangledown$] If $c=1$, the coefficient of $\boxed{u_x^2}$ is $-k+\frac{2}{3}w$. We therefore take $c=-1$.
    \end{itemize}
     \item The coefficient of $\boxed{u_x^2}$ gives $-2 h_3{}^{(0,1)}(x,w)-k\Rightarrow h_3(x,w)=k_2(x)-\frac{k w}{2}$.
     \item Equation $D_t(P)=0$ takes the form $\frac{1}{6} k_1'(x) \left((v-u) (3 k+u-v)+6 k_1(x)\right)+k_1{}^{(3)}(x)$. Therefore, $k_1$ is a constant.
\end{itemize}

Finally, we have $D_t(P)=0$ and
\begin{equation}
    v_x+u_x=-\frac{1}{6} (u-v)^2-k (u-v)+k_1
\end{equation}

After substitution $u\to6u,\:k_1\to6\alpha$, we obtain the form \eqref{B-pKdV}.

\subsection{Korteweg de Vries equation}

\begin{equation}
    u_t=u_{xxx}+u_xu
\end{equation}

The potKdV equation belongs to case A in \eqref{Appcond}, so its B\"acklund transformation has the form

\begin{equation}
    v_x=c(x,u,v)u_x+h(x,u,v)
\end{equation}

\begin{itemize}
  \item The coefficient of $\boxed{u_{xxx}}$ is $D\;c(x,u,v)=0$, so, by our convention, $c$ is constant.
  
  \item The coefficient of $\boxed{u_{xx}u_x}$ is $(\partial_u+c\partial_v)^2h=0\Rightarrow h=h_1(x,v-cu)\cdot(c u+v)+h_2(x,v-c u)$.
  \item The coefficient of $\boxed{u_x^2}$ is $c \left(12 c h_1(x,w) \partial_wh_1(x,w)+c-1\right)=0\:$, where $,w=v-cu\Rightarrow $. Solving gives \\$h_1(x,w)=\pm \sqrt{\frac{(1-c) w}{6 c}+k(x)}$; the sign corresponds to the symmetry $t\to-t;x\to-x$.
  \item The coefficient of $\boxed{u_{xx}}$ yields an algebraic equation for $h_2\Rightarrow$
  \begin{equation*}
     h_2(x,w)=\frac{36 c k'(x)+(c+1) w \sqrt{6 \left(\frac{1}{c}-1\right) w+36 k(x)}}{6 (c-1)}
  \end{equation*}
  \item The coefficient of $\boxed{u_{xx}}$ has the form $c(c+1)(F(x,v,u))$, where $F\neq0$. Thus, $c=-1$.
  \item The coefficient of $\boxed{u_x}$ after the substitution $u=-v$ gives $3 \left(v k''(x)+k^{(4)}(x)\right)+\frac{9}{2} k'(x)^2\Rightarrow$ k is independent of x.
\end{itemize}

Finally, we have $D_t(V)=0$, and the general B\"acklund transformation is
\begin{equation}
    v_x+u_x=\pm(u-v) \sqrt{k+\frac{1}{3} (-u-v)}
\end{equation}

After substituton $u\to-3u,\:k\to\alpha$, we obtain the form \eqref{B-KdV}.

\subsection{Modified KdV}
Consider the modified KdV (mKdV) equation
\begin{equation}
    u_t=u_{xxx}+u_xu^2
\end{equation}

The first two steps are the same as for KdV. 

\begin{equation}
    v_x-c u_x= h_1(x,v-c u) v+h_2(x,v-c u)
\end{equation}

\begin{itemize}
    \item The coefficient of $\boxed{u_x^2u}$ after the substitution $v=w+cu$ gives $2 c^2 \left(c^2-1\right)\Rightarrow c=\pm1$. The sign corresponds to the symmetry $v\to-v$, so we choose $c=1$.
     \item The coefficient of $\boxed{u_{xx}u}$ after the substitution $v=w+cu$ gives $ \left(3 h_1(x,w) \partial_wh_1(x,w)+2 w\right),\:w=v-u\Rightarrow$ $h_1(x,w)=\sqrt{k(x)-\frac{2 w^2}{3}}$.
      \item The coefficient of $\boxed{u_{xx}u}$ after the substitution $v=w+cu$ yields an algebraic equation for $h_2\Rightarrow h_2(x,w)=\frac{9 k'(x)-2 w^2 \sqrt{9 k(x)-6 w^2}}{12 w}$.
      \item The coefficient of $\boxed{u_xu}$ after the substitution $v=w+cu$ gives $-\frac{3 \left(w^2-3 k(x)\right) k'(x)}{2 w^3}\Rightarrow k$ is independent of x.
\end{itemize}

Finally, we have $D_t(P)=0$ and

\begin{equation}
    v_x\pm u_x=\frac{1}{6} (v\mp u) \sqrt{9 k-6 (v\pm u)^2}
\end{equation}
After substitution $u\to i\sqrt{6}u,\:x\to-x,\:t\to-t,\:k\to 4\alpha$, we obtain the form \eqref{B-mKdV}.

\subsection{The exponential Calogero--Degasperis equation}
Consider the exponential Calogero--Degasperis equation
\begin{equation}
    u_t = u_{xxx} - \frac{1}{2}u_x^3 -\frac{3}{2}(c_1e^{2u} + c_2e^{-2u})u_x
\end{equation}

The first step is the same as before:

\begin{equation}
    v_x-c u_x= h(x,v,u)
\end{equation}

\begin{itemize}
    \item The coefficient of $\boxed{u_{xx}u_x}$ gives $-3c^2h+3(\partial_u+c\partial_v)^2h\Rightarrow$ $h=e^v h_1(x,v-c u)+e^{-v} h_2(x,v-c u)$.
    \item The coefficient of $\boxed{u_{xx}u_x^2}$ is $-\frac{3}{2}c ( c^2-1)$.
\end{itemize}

For the minus sign, $c=1$.

\begin{itemize}
    \item The coefficient of $\boxed{u_{xx}}$ after the substitution $u=v-w$ has the form $a_2\exp{2v}+a_1\exp{v}+a_0+a_{-1}\exp{(-v)}+a_{-2}\exp{(-2v)}$, where $a_i$ are functions of $w,x$. Therefore, $a_1=3\partial_xh_1=0,a_{-1}=3\partial_xh_2=0$. Thus, $h_1,h_2$ are independent of $x$.
    \item Now $a_{2}$ equals $3 \left(c_1 \left(e^{-2 w}-1\right)+2 h_1(w) h_1'(w)+h_1(w){}^2\right)$ and must vanish. The resulting equation has two solutions, related by the symmetry $x\to-x,t\to-t$. We therefore take $h_1=\sqrt{c_1 e^{-2 w}+c_2+k e^{-w}}$.
   \item We have $a_0$ equal to $$\frac{e^w h_2(w) \left(2 c_1 e^w+k\right) \sqrt{c_1 e^{-2 w}+c_1+k e^{-w}}}{c_1 e^{2 w}+c_1+k e^w}-2
   \sqrt{c_1 e^{-2 w}+c_1+k e^{-w}} h_2'(w),$$ and the equation $a_0=0$ has the solution $h_2=k_2 \sqrt{c_1 e^{2 w}+c_1+k e^w}$.
   \item After substitution, $a_{-2}$ equals $\frac{3}{2} \left(e^{2 w}-1\right) \left(c_2-c_1 k_2^2\right)$, so $k_2=\pm\sqrt{\frac{c_2}{c_1}}$. For these values of $k_2$, the equation $D_tP=0$ holds identically.
   
\end{itemize}

Thus, for $c=1$ and nonzero constants $c_1,c_2$, we obtain the general B\"acklund transformation

\begin{equation}
    v_x-u_x=e^v \sqrt{c_1 \left(k e^{u-v}+e^{-2 (v-u)}+1\right)}\pm e^{-v} \sqrt{c_2 \left(k e^{v-u}+e^{2 (v-u)}+1\right)}
\end{equation}
and the same equation with $x\to-x$. After changing the variables and notation, we obtain the form \eqref{B-expCD1}.

For the plus sign, $c=-1$.

\begin{itemize}
    \item The coefficient of $\boxed{u_{xx}}$ after the substitution $u=w-v$ has the same form $a_2\exp{2v}+a_1\exp{v}+a_0+a_{-1}\exp{(-v)}+a_{-2}\exp{(-2v)}$, where $a_i$ are functions of $w,x$. Similarly, we find that $h_1,h_2$ are independent of $x$ and that $h_1=\pm\sqrt{c_2 e^{-2 w}+c_1+k e^{-w}}$, where the sign corresponds to the symmetry $x\to-x,t\to-t$.
    \item From $a_{0}$, we obtain $h_2=k_2 \sqrt{c_1 e^{2 w}+c_2+k e^w}$.
    \item Finally, after substitution, $a_{-2}$ gives $-\frac{3}{2} \left(k_2^2-1\right) \left(c_2-c_1 e^{2 w}\right)\Rightarrow k_2=\pm1$. For these values of $k_2$, the equation $D_tP=0$ holds identically.
 
Thus, for $c=-1$ and nonzero constants $c_1,c_2$, we obtain the general B\"acklund transformation
\end{itemize}
\begin{equation}
    v_x+u_x=e^{-v} \sqrt{c_1 e^{2 (u+v)}+c_2+k e^{u+v}}\pm e^v \sqrt{c_2 e^{-2 (u+v)}+c_1+k e^{-u-v}}
\end{equation}
and the same equation with $x\to-x$. After changing the variables and notation, we obtain the form \eqref{B-expCD2}.

Degenerate case 

\begin{remark}
    The coefficient $a_0$ can vanish in both cases $c=\pm1$ if $c_1=c_2=0$.
\end{remark}

In this case, the condition involving $D_tP$ holds identically without restrictions on $k_2$. For $c=1$, we have

\begin{equation}
    v_x- u_x=k_1 e^v \sqrt{e^{u-v}}+k_2 e^{-v} \sqrt{e^{v-u}}
\end{equation}
and for $c=-1$, we have
\begin{equation}
   v_x+u_x= k_1 e^{-v} \sqrt{e^{u+v}}+k_2 e^v \sqrt{e^{-u-v}}
\end{equation}

where $k_1,k_2$ are arbitrary constants. Thus, we obtain \eqref{B-expCD3}.

\subsection{The Krichever--Novikov equation}
Consider the Krichever--Novikov (KN) equation
\begin{equation}
    u_t=u_{xxx}-\frac{3}{2}\frac{u_{xx}^2}{u_x}+\frac{q(u)}{u_x}
\end{equation}

The KN equation belongs to case C in the list above:

\begin{equation}
    \log(v_x)-c(x,u,v)\log(u_x)=p(x,u,v)
\end{equation}

\begin{itemize}
    \item The coefficient of $\boxed{v_{xxx}}$ is $D_x(c)$, so $c$ is constant.
    \item The coefficient of $\boxed{v_{xx}^3}$ gives $\sim (1-c^2)\Rightarrow c=\pm 1$.
    \item The coefficient of $\boxed{v_{xx}^2}$ under the assumption $c^2=1$ gives $p_x\Rightarrow p$ does not depend on $x$.
    \item For $c=-1$, consider the coefficients of $\boxed{v_{xx}}$ and $\boxed{v_x^4}$. The first is $\sim (2q(u)-3p^2_v+6pp_{vv})$ and the second is $\sim (2q(v)-3p^2_u+6pp_{uu})$. The respective solutions are $$p=g(u)+h(u)v+\frac{(3 h(u)^2-2q(u))v^2}{12g(u)}, \quad  p=g(v)+h(v)u+\frac{(3 h(v)^2-2q(v))u^2}{12g(v)}.$$ We therefore deduce that $p$ is a quadratic polynomial in $v$ and $u$.
    \item After substituting p as a polynomial, the condition involving $D_t(B)$ becomes a polynomial equation in the constants. Solving it gives equations for the coefficients of the B\"acklund transformation for the KN equation:
\end{itemize}

\begin{equation}
v_xu_x=a_1v^2u^2+a_2vu(v+u)+a_3(v^2+u^2)+a_5vu+a_6(v+u)+a_9,
\end{equation}

where the coefficients of the B\"acklund transformation satisfy \eqref{KN-B-coeff}.

When $c=1$, the coefficient of $\boxed{v_{xx}}$ gives $p(u,v)^2=\frac{q(v)}{q(u)}$. One can then check that this cannot satisfy the equation $D_t(V)=0$.

\subsection{Schwarzian KdV}

Consider the Schwarzian KdV equation

\begin{equation}
    u_t=u_{xxx}-\frac{3}{2}\frac{u_{xx}^2}{u_x}
\end{equation}

The Schwarzian KdV equation belongs to case C in the list above:

\begin{equation}
    \log(v_x)-c(x,u,v)\log(u_x)=p(x,u,v)
\end{equation}

This is the Krichever--Novikov equation with $q(u)=0$. The first five steps do not use the form of $q(u)$, so the B\"acklund transformation for Schwarzian KdV has the same form, with $p(u,v)$ a quadratic polynomial in $v$ and $u$. Solving the resulting system of equations for the constants gives \eqref{BB-ShKdV}. This is therefore the general form of the B\"acklund transformation.

\subsection{The potential Gardner equation}
Consider the potential Gardner equation

\begin{equation}
    u_t = u_{xxx} - \frac{3u_{xx}^2}{4u_x} + a_1 u_x^{3/2} -3 a_2^2 u_x^2, \quad a_1 \neq 0 \text{ or } a_2 \neq 0,
\end{equation}

It belongs to case B in the list.

\begin{equation}
    \sqrt{v_x}=c(x,u,v)\sqrt{u_x}+h(x,u,v)
\end{equation}

\begin{itemize}
    \item The coefficient of $\boxed{v_{xxx}}$ gives $D_x(c(x,v,u))=0$, so $c(x,u,v)$ is constant.
    \item The coefficient of $\boxed{v_{xx}^2}$ gives $-\frac{3 \left(c^2 \partial_v h+\partial_u h\right)}{2 c^2 v}\Rightarrow h=h(x,v-c^2u)$.
    \item The coefficient of $\boxed{v_{xx}v_x^2}$ is $-\frac{3}{4} \left(-2 h \left((\partial_yh)^2+h \partial_y^2h\right)+\frac{a_1}{4 c}+\frac{a_1}{4}\right)$, where $y=v-c^2u$. Thus, we obtain an equation for $\partial_y^2h$.
    \item The coefficient of $\boxed{v_{xx}}$ at $v_x=0$, after substitution of $\partial_y^2h$, is
    \begin{equation}\label{eq1}
        h \left(3 \partial_x\partial_yh+\frac{3}{8} a_1 \left(\frac{1}{c}-1\right)\right)+3 \partial_xh \partial_yh
    \end{equation}
    Thus, we obtain an equation for $\partial_x\partial_yh$.
    \item After substituting the two previously obtained equations, the coefficient of $\boxed{v_1}$ is $-12\partial_xh(\partial_yh)^2$. Therefore, $h$ is independent of x. From $\ref{eq1}$, we see that either $c=1$ or $a_1=0$.
     \item For $c=1$, the equation for $\partial_y^2h$ is sufficient for $D_t(V)=0$. However, the solution of this ODE is cumbersome, and we do not consider it here.
       \item[$\vartriangle$] For $a_1=0$, the equation for $\partial_y^2h$ gives $h=k_2\sqrt{2y-k_1}$. The compatibility condition $D_tP=0$ then holds.
\end{itemize}

Thus, the general B\"acklund transformation for $a_1=0$ is

\begin{equation}
    \sqrt{v_x}+\sqrt{u_x}=a_2\sqrt{(v-u)^2-k_1}
\end{equation}

\section{Volterra-chains}

The following calculations outline the explicit solution of the consistency conditions $D_\tau(V)=0$, where $V$ is as in \eqref{Backl-form} and is already known from the calculations above. The unknown function is $h(v,u,w)$, where $u_\tau=h(v,u,w)$. A general scheme for solving such conditions is described in \cite{Garif_2021}. The first step is always to differentiate $D_\tau(V)$ with respect to $u_2$. We use the notation $\log(\partial_{u_2}h(u_2,v,u))=F$ and shift the equation one step in the inverse direction.
\begin{remark}
    We assume the absence of pseudoconstants, i.e., that the only solutions of the  equations $D_x(c)=0 \quad (T-1)c=0$   are constants.
\end{remark}

\subsection{Potential KdV}
Consider the B\"acklund transformation for potential KdV

\begin{equation}
    v_x+u_x=-(u-v)^2-k (u-v)+\al
\end{equation}

For the compatibility condition $D_\tau(V)=0$, the general scheme requires us to consider $\partial_{u_2}  D_\tau(B)$ and write $D_x(\log(\partial_{u_2} h))=F$. The coefficient of $u_x$ is then $-\partial_wF(v,u,w)+\partial_uF(v,u,w)-\partial_vF(v,u,w)$, so the dependence is of the form

$$F\left(w-v,u+v\right)$$

After the substitution

$w\to l+v,u\to \eta-v$, the equation becomes polynomial in $\eta$. The coefficient of $v^2$ is $-8 \partial_\eta F(l,\eta)$, so $F$ is independent of $\eta$. Then $D_\tau(V)$ is linear in $u$, with coefficient $-4 \left(-(k-\eta+l) F'(l-\eta)-2\right)$. Hence, $F=-2 \log (k-\eta+l)+c_1$ and $h=-\frac{e^{c_1}}{k-v+w}+f(u,w)$. Substitution shows that $f$ is constant. We therefore obtain the general form of the three-point symmetry:
\begin{equation}
    u_\tau=-\frac{e^{c_1}}{k-v+w}+c_2
\end{equation}
\subsection{Korteweg de Vries equation}
Consider the B\"acklund transformation for the KdV equation
\begin{equation}\label{AB-KdV}
    v_x+ u_x= (v-u) \sqrt{ v+u+\al}
\end{equation}
Again, consider the compatibility condition $D_\tau(V)=0$ and write $D_x(\log(\partial_{u_2} h))=F$. The coefficient of $u_x$ is $-\partial_wF(v,u,w)+\partial_uF(v,u,w)-\partial_vF(v,u,w)$, so $F$ has the dependence indicated by $F(v+u,w+u)$. Introducing $v+u=l,u+w=\eta$ gives the equation $D_\tau(V)=Q+Gu$, where $$G=-2 \sqrt{\al+\eta} \partial_\eta F(l,\eta)+2 \sqrt{\al+l}
   \partial_lF(l,\eta)-\frac{1}{\sqrt{\al+\eta}}-\frac{1}{\sqrt{\al+l}}=0$$
   $$Q=
   \eta \sqrt{\al+\eta} \partial _\eta F(l,\eta)-l \sqrt{\al+l}
  \partial_lF(l,\eta)-\frac{2\al+\eta }{2\sqrt{\al+\eta}}+\frac{2\al+3l}{2\sqrt{\al+l}}=0$$
Solving this system for $\partial_lF(l,u)$ and integrating the resulting ordinary differential equation gives $$F(l,u)= \frac{1}{2} \left(4 \log
   \left(\sqrt{\al+\eta}+\sqrt{\al+l}\right)+\log
   (\al+l)\right)+c_1(u).$$ Substituting this into $Q$ yields another ordinary differential equation, whose solution gives $$\exp(F)=\frac{e^{-c_1} \sqrt{\al+\eta}}{\sqrt{\al+l}
   \left(\sqrt{\al+\eta}+\sqrt{\al+l}\right)^2}\Rightarrow h=-\frac{2 e^{-\text{c1}}
   \sqrt{w+u+\al}}{\sqrt{w+u+\al}+\sqrt{u+v+\al}}+P(u,w)$$
The B\"acklund transformation has the symmetry $n\to -n $, so $$ h=-\frac{2 e^{-\text{c1}}
   \sqrt{v+u+\al}}{\sqrt{w+u+\al}+\sqrt{u+v+\al}}+\hat P(u,v)$$
Differentiating both formulas, subtracting, and integrating gives
   $$h(v,u,w)=\frac{e^{-\text{c1}}
   \left(\sqrt{\al+u+v}-\sqrt{\al+u+w}\right)}{\sqrt{\al+u+w}+\sqrt{\al+u+v}}+S(u)$$
After substitution, the coefficient in $D_\tau(V)$ of $u_x$ is proportional to $S'(u)-S'(v)$. Thus, $S(u)$ is linear in $u$. The condition involving $D_\tau(V)$ then implies that this linear function vanishes. We therefore obtain the general discrete symmetry for KdV:
   \begin{equation}
       u_\tau=\frac{\left(\sqrt{\al+u+v}-\sqrt{\al+u+w}\right)}{\sqrt{\al+u+w}+\sqrt{\al+u+v}}
   \end{equation}
\subsection{Modified KdV}
Consider the B\"acklund transformation for mKdV

\begin{equation}\label{AB-mKdV}
    v_x+ u_x= (v-u) \sqrt{ (v+ u)^2+\alpha}
\end{equation}

The first two steps are identical to those for KdV. Again, we obtain two equations for $F(u+v,w+u)=F(l,\eta)$:

$2 \sqrt{\al+\eta^2} \partial_\eta F(l\eta)-2\sqrt{\al+l^2}\partial_l F(l,\eta)-\frac{2 \eta}{\sqrt{\al+\eta^2}}-\frac{2l}{\sqrt{\al+l^2}}$

$$-\eta \sqrt{\al+\eta^2} \partial_\eta F(l,\eta)+l \sqrt{\al+l^2}  \partial _l F(l,\eta)-\frac{\al}{\sqrt{a+\eta^2}}+\frac{2l^2+\al}{\sqrt{\al+l^2}}$$
Solving these equations for one derivative and integrating gives
   $$F(l,\eta)=-\log\left(\sqrt{\al+\eta^2} \sqrt{\al+l^2}+\al+\eta l\right)+\frac{1}{2} \log\left(\al+\eta^2\right)+f(l)$$
We then determine $f$ from the same equations:
 $$f(l)=-\frac{1}{2} \log \left(\al+l^2\right)+c_1$$
Thus, $h$ has the form

   $$h(v,u,w)=\frac{e^{c_1} \sqrt{\al+\eta^2} \left(\sqrt{\al+\eta^2}-\sqrt{\al+l^2}\right)}{\al
   (\eta-l)}+P(u,w)$$
where $l=v+u,\eta=u+w=$

The equation has the symmetry $V(-n)=-V(n)$, so, as in the KdV case, we obtain

$$h(v,u,w)=\frac{e^{c_2} \sqrt{\al+l^2} \left(\sqrt{\al+l^2}-\sqrt{\al+\eta^2}\right)}{\al
   (l-\eta)}+Q(v,u)$$

and therefore

   $$h(v,u,w)=-\frac{e^{c_1} \left(-\sqrt{\al+\eta^2} \sqrt{\al+l^2}+\al+\eta l\right)}{\al (\eta-l)}+S(u)$$

Substitution gives $S=0$. Thus, the general form of the discrete symmetry is
   \begin{equation}
       u_\tau=-\frac{ \left(-\sqrt{\al+\eta^2} \sqrt{\al+l^2}+\al+\eta l\right)}{\al (\eta-l)}
   \end{equation}
 
\subsection{The exponential Calogero--Degasperis equation}
Consider the B\"acklund transformation for the exponential Calogero--Degasperis equation
\begin{equation}\label{AB-expCD1}
    v_x+u_x=(e^{-v}+e^{-u})
 \sqrt{c_1e^{2(v+u)}+2\al e^{v+u}+c_2}
\end{equation}
The first step is the same as before and gives $F=F(v+u,u+w)$. Introducing $v+u=l,u+w=\eta$ then makes the expression $D_\tau(V)$ polynomial in $\exp(u_1)$. We use the notation $\sqrt{e^l \left(c_1 e^l+2 \al\right)+c_2}=r(u,v),\sqrt{e^\eta \left(c_1 e^\eta+2\al\right)+c_2}=r(u,w)$. As before, each coefficient must vanish; these coefficients are $2 r(u,w) e^{-\eta} \partial_\eta F(l,\eta)-2 r(u,v) e^{-l} \partial_lF(l,\eta)-\frac{2c_1e^\eta+2\al}{r(u,w)}+\frac{2c_2 e^{-l}+2\al}{r(u,v)}$ and $-r(u,w) \partial_\eta F(l,\eta)+r(u,v) \partial_l F(l,\eta)-\frac{2 \al e^\eta+c_2}{2r(u,w)}+\frac{c_1 e^{2 l}+2\al e^l}{r(u,v)}$.
Solving this linear ODE system gives
   
\[
\begin{aligned}
F(l,\eta)= \frac12\bigl[-2\log(2c_1e^{\eta+l}+2r(u,w)r(u,v) +2\alpha e^\eta+2\alpha e^l+2c_2)+\\\log(c_1e^{2l}+2\alpha e^l+c_2)+\log(c_1e^{2\eta}+2\alpha e^\eta+c_2)+2l\bigr]+k_1.
\end{aligned}
\]
where $k_1$ is a constant. Therefore,

   $$h(v,u,w)=-\frac{2 e^{c1} r(u,w) (r(u,w)-r(u,v))}{\left(4\al^2-4 c_1 c_2\right)
   \left(e^\eta-e^l\right)}+p(u,v)$$

Differentiating the consistency condition involving $D_\tau(V)$ with respect to $u_{-1}$ and solving analogous equations gives

$$h(v,u,w)=\frac{2 e^{c1} r(u,w) (r(u,w)-r(u,v))}{\left(4\al^2-4 c_1 c_2 \right)
   \left(e^\eta-e^l\right)}+\hat p(v,u)$$

where $p,\hat p$ are unknown functions. Combining the formulas above, we obtain
   $$h(v,u,w)=-\frac{e^{-u+c_1} \left(2\al e^u \left(e^{w}+e^{v}\right)+2 c_1 e^{w+2u+v}+2 c_2-2
   r(u,w) r(u,v)\right)}{2 \left(e^{v}-e^{w}\right)}+p(u)$$
Substitution shows that $p(u)=0$. We therefore obtain the general form of the discrete symmetry for the B\"acklund transformation \eqref{B-expCD1}:

   \begin{equation}
       u_\tau=-\frac{e^{-u} \left(\al e^u \left(e^{w}+e^{v}\right)+ c_1 e^{w+2u+v}+c_2-
   r(u,w) r(u,v)\right)}{ \left(e^{v}-e^{w}\right)}
   \end{equation}

In the same way, one can obtain discrete symmetries for \eqref{B-expCD2} and \eqref{B-expCD3} and verify that the symmetries obtained from the quad equations are indeed the general symmetries.

\subsection{The Krichever--Novikov equation}
Consider the B\"acklund transformation for the Krichever-Novikov equation

\begin{equation}\label{AB-KN}
v_xu_x=a_1v^2u^2+a_2vu(v+u)+a_3(v^2+u^2)+a_5vu+a_6(v+u)+a_9,
\end{equation}

The expression $D_\tau(V)$ is polynomial in $u_x$, and its free term contains only $\partial_uF(v,u,w)$. Solving the resulting differential equation gives $F=\log(P)+\hat P(w,v)$, where $P$ is a polynomial in all three variables. Now $D_\tau(V)$ is polynomial in $v$. Solving the linear differential system for $\hat P(w,v)$ gives $\hat P(w,v)=w-v+k_1$ and hence $$h=\frac{P(v,u,u_{-1})}{v-u_{-1}}+\zeta(u,u_{-1}),$$ where $P$ is a polynomial and $\zeta$ is an unknown function. Differentiating the consistency condition involving $D_\tau(V)$ with respect to $u_{-1}$ and solving analogous equations gives $$h=\frac{\widetilde P(v,u,w)}{v-w}+\widetilde \zeta(v,u).$$ Combining these two expressions and imposing the condition on $D_\tau(B)$ to determine the remaining function of $u$ yields the following three-point discrete symmetry:

\begin{equation}
    u_\tau=\frac{R(v,u,w)}{v-w}
\end{equation}

where $R(v,u,w)=2vw(a_1u^2+a_2u+a_3)+(v+w)(a_2u^2+a_5u+a_6)+2a_3u^2+2a_6u+2a_9$.
The coefficients of the B\"acklund transformation and the discrete symmetry satisfy \eqref{KN-B-coeff}.

\subsection{Schwarzian KdV}
Consider the B\"acklund transformation for Schwarzian KdV
\begin{equation}
   v_xu_x=\al(u-v)^2
\end{equation}

The first step is the same: differentiate $D_\tau(V)$ and use the notation $\log(\partial_{u_2}h(u_2,v,u))=F$. Then $D_\tau(V)$ is polynomial in $u_x$, and its free term is proportional to $\left((u-w) \partial_uF(v,u,w)-2\right)$. Therefore,
\begin{equation}
    F(u,v,w)=2\log(w-u)+k(w,v)
\end{equation}
After substitution, the coefficient of $u_x$ gives $k(u,w)=-2\log(w-v)+\hat k(u)$ and then implies that $k(u)$ is constant. Differentiating the consistency condition involving $D_\tau(V)$ with respect to $u_{-1}$ and solving analogous equations gives
\begin{equation}
    h(v,u,w)=k\left(\frac{(u-v)^2}{v-w}+v\right)+p(u)
\end{equation}
where p is an unknown function. The consistency condition gives $p=k_2 u$. Changing the notation then yields the general form of the discrete symmetry:
\begin{equation}\label{V-ShKdV}
    u_\tau=k_1\frac{(u-v)(u-w)}{w-v}+k_2u
\end{equation}

\subsection{The potential Gardner equation}
Consider the B\"acklund transformation for the potential Gardner equation \eqref{pGar} when $a_1=0$:
\begin{equation}
    \sqrt{v_x}+\sqrt{u_x}=a_2\sqrt{(v-u)^2-\al}
\end{equation}

The first two steps are the same as for KdV, except that $F$ depends on $v-u=l,u-w=\eta$. After these steps, $D_\tau(V)$ is polynomial in $\sqrt{v_x}$, which gives the following equations:

$$\sqrt{\eta^2-\al} \partial_\eta F(l,\eta)+\sqrt{l^2-\al}
   \partial_lF(l,\eta)-\frac{\eta}{\sqrt{\eta^2-\al}}+\frac
   {l}{\sqrt{l^2-\al}}=0$$

\begin{equation}
    \frac{1}{2} \left(\left(-2 \sqrt{\eta^2-\al}
   \sqrt{l^2-\al}-\al+l^2\right)
   \partial_lF(l,\eta)+\left(\al-\eta^2\right)
   \partial_\eta F(l,\eta)\right)-\frac{l
   \sqrt{\eta^2-\al}}{\sqrt{l^2-\al}}+\eta+l=0
\end{equation}
Solving this system gives

   $$F(l,\eta)=-\log \left(\sqrt{\eta^2-\al}
   \sqrt{l^2-\al}+\al-\eta l\right)+\frac{1}{2} \log
   \left(\eta^2-\al\right)-\frac{1}{2} \log
   \left(\al-l^2\right)+c_1$$
from which we obtain

   $$h(v,u,w)=\frac{e^{c_1} \sqrt{\eta^2-\al}
   \left(\sqrt{\eta^2-\al}+\sqrt{l^2-\al}\right)}{\al
   (l-\eta)}+p(u,u_{-1})$$

Differentiating the consistency condition involving $D_\tau(V)$ with respect to $u_{-1}$ and solving analogous equations gives

   $$h(v,u,w)=\frac{e^{k} \sqrt{l^2-\al}
   \left(\sqrt{\eta^2-\al}+\sqrt{l^2-\al}\right)}{\al
   \eta-\al l}+p(v,u) $$

from which we obtain
   \begin{equation}
       h(v,u,w)=e^{k}\frac{(u-w)(u-v)-\al-\sqrt{l^2-\al}\sqrt{\eta^2-\al}}{v-w}+\hat k(u)
   \end{equation}
Finally, the consistency condition implies that $k(u)$ is constant. Thus, we obtain the general form of the three-point discrete symmetry:
\begin{equation}\label{V-pGar}
    u_\tau=k_1\frac{(u-w)(u-v)-\al-\sqrt{l^2-\al}\sqrt{\eta^2-\al}}{v-w}+k_2
\end{equation}

where $k_1,k_2$ are arbitrary constants.


\begin{thebibliography}{99}

\bibitem{Hasegawa_Tappert} A. Hasegawa, F. Tappert, Transmission of stationary nonlinear optical
pulses in dispersive dielectric fibers. I. Anomalous dispersion, Appl. Phys. Lett. 23 (1973) 142--144.

\bibitem{Ustinov} A. V. Ustinov, Solitons in Josephson junctions, Physica D 123 (1998) 315--329.

\bibitem{MSY_1987} A. V. Mikhailov, A. B. Shabat, R. I. Yamilov, The symmetry approach to the
classification of non-linear equations. Complete lists of integrable systems, Russian Math.
Surveys 42:4 (1987) 1--63.

\bibitem{Mikhailov_Shabat_Sokolov} A. V. Mikhailov, A. B. Shabat, V. V. Sokolov, The symmetry
approach to classification of integrable equations, in: V. E. Zakharov (ed.), What is
Integrability?, Springer-Verlag, Berlin, 1991, pp.~115--184.

\bibitem{Sokolov_book} V. V. Sokolov, Algebraic Structures in Integrability, World Scientific,
Singapore, 2020.

\bibitem{Svinolupov_Sokolov_1982} S. I. Svinolupov, V. V. Sokolov, Evolution equations with
nontrivial conservation laws, Funct. Anal. Appl. 16:4 (1982) 317--319.

\bibitem{Meshkov_Sokolov} A. G. Meshkov, V. V. Sokolov, Integrable evolution equations with a
constant separant, Ufa Math. J. 4:3 (2012) 104--154.

\bibitem{Yamilov_1983} R. I. Yamilov, Classification of discrete evolution equations, Uspekhi
Mat. Nauk 38:6 (1983) 155--156 (in Russian).

\bibitem{Yamilov} R. Yamilov, Symmetries as integrability criteria for differential difference
equations, J. Phys. A: Math. Gen. 39 (2006) R541--R623.

\bibitem{Levi_1981} D. Levi, Nonlinear differential-difference equations as Bäcklund
transformations, J. Phys. A: Math. Gen. 14 (1981) 1083--1098.

\bibitem{Shabat_Yamilov_1990} A. B. Shabat, R. I. Yamilov, Symmetries of nonlinear lattices,
Algebra i Analiz 2:2 (1990) 183--208; English transl.: Leningrad Math. J. 2:2 (1991) 377--400.

\bibitem{Veselov_Shabat} A. P. Veselov, A. B. Shabat, Dressing chains and spectral theory of the
Schrödinger operator, Funct. Anal. Appl. 27:2 (1993) 81--96.

\bibitem{Wahlquist_Estabrook} H. D. Wahlquist, F. B. Estabrook, Bäcklund transformation for
solutions of the Korteweg--de Vries equation, Phys. Rev. Lett. 31 (1973) 1386--1390.

\bibitem{Rogers_Schief} C. Rogers, W. K. Schief, Bäcklund and Darboux Transformations: Geometry
and Modern Applications in Soliton Theory, Cambridge University Press, Cambridge, 2002.

\bibitem{Hirota_1977} R. Hirota, Nonlinear partial difference equations. I. A difference analogue
of the Korteweg--de Vries equation, J. Phys. Soc. Japan 43 (1977) 1424--1433.

\bibitem{NQCL_1984} G. R. W. Quispel, F. W. Nijhoff, H. W. Capel, J. van der Linden, Linear
integral equations and nonlinear difference-difference equations, Physica A 125 (1984) 344--380.

\bibitem{Nijhoff_Capel} F. Nijhoff, H. Capel, The discrete Korteweg--de Vries equation, Acta
Appl. Math. 39 (1995) 133--158.

\bibitem{HJN_book} J. Hietarinta, N. Joshi, F. W. Nijhoff, Discrete Systems and Integrability,
Cambridge University Press, Cambridge, 2016.

\bibitem{Nijhoff_2002} F. W. Nijhoff, Lax pair for the Adler (lattice Krichever--Novikov) system,
Phys. Lett. A 297 (2002) 49--58.

\bibitem{Bobenko_Suris_2002} A. I. Bobenko, Yu. B. Suris, Integrable systems on quad-graphs, Int.
Math. Res. Not. 2002:11 (2002) 573--611.

\bibitem{ABS} V. E. Adler, A. I. Bobenko, Yu. B. Suris, Classification of integrable equations on
quad-graphs. The consistency approach, Commun. Math. Phys. 233 (2003) 513--543.

\bibitem{ABS_2009} V. E. Adler, A. I. Bobenko, Yu. B. Suris, Discrete nonlinear hyperbolic
equations. Classification of integrable cases, Funct. Anal. Appl. 43:1 (2009) 3--17.

\bibitem{Krichever_Novikov} I. M. Krichever, S. P. Novikov, Holomorphic bundles over algebraic
curves and nonlinear equations, Russian Math. Surveys 35:6 (1980) 53--80.

\bibitem{Adler_1998} V. E. Adler, Bäcklund transformation for the Krichever--Novikov equation,
Int. Math. Res. Not. 1998:1 (1998) 1--4.

\bibitem{Adler_Suris_Q4} V. E. Adler, Yu. B. Suris, $Q_4$: integrable master equation related to
an elliptic curve, Int. Math. Res. Not. 2004:47 (2004) 2523--2553.

\bibitem{Atkinson} J. Atkinson, M. Nieszporski, Multi-quadratic quad equations: integrable cases
from a factorized-discriminant hypothesis, Int. Math. Res. Not. 2014:15 (2014) 4215--4240.

\bibitem{Kels} A. P. Kels, Integrable quad equations derived from the quantum Yang--Baxter
equation, Lett. Math. Phys. 110 (2020) 1477--1557.

\bibitem{Atkinson_2008} J. Atkinson, Bäcklund transformations for integrable lattice equations,
J. Phys. A: Math. Theor. 41 (2008) 135202.

\bibitem{Rasin_Hydon} O. G. Rasin, P. E. Hydon, Symmetries of integrable difference equations on
the quad-graph, Stud. Appl. Math. 119 (2007) 253--269.

\bibitem{Tongas_Tsoubelis_Xenitidis} A. Tongas, D. Tsoubelis, P. Xenitidis, Affine linear and
$D_4$ symmetric lattice equations: symmetry analysis and reductions, J. Phys. A: Math. Theor. 40
(2007) 13353--13384.

\bibitem{LPSY_2008} D. Levi, M. Petrera, C. Scimiterna, R. Yamilov, On Miura transformations and
Volterra-type equations associated with the Adler--Bobenko--Suris equations, SIGMA 4 (2008) 077.

\bibitem{Tsoubelis_Xenitidis} D. Tsoubelis, P. Xenitidis, Continuous symmetric reductions of the
Adler--Bobenko--Suris equations, J. Phys. A: Math. Theor. 42 (2009) 165203.

\bibitem{Xenitidis_2009} P. Xenitidis, Integrability and symmetries of difference equations: the
Adler--Bobenko--Suris case, in: Proceedings of the 4th Workshop ``Group Analysis of Differential
Equations and Integrable Systems'', Protaras, Cyprus, 2008; arXiv:0902.3954.

\bibitem{Verosky} J. M. Verosky, Negative powers of Olver recursion operators, J. Math. Phys. 32
(1991) 1733--1736.

\bibitem{Adler_2024} V. E. Adler, 3D consistency of negative flows, Theor. Math. Phys. 221 (2024)
1836--1851.

\bibitem{Kolesnikov_2025} M. P. Kolesnikov, The negative symmetry classification problem, Theor.
Math. Phys. 224 (2025) 1398--1413.

\bibitem{AK_2022} V. E. Adler, M. P. Kolesnikov, Non-Abelian Toda lattice and analogs of
Painlevé III equation, J. Math. Phys. 63:10 (2022) 103504.

\bibitem{AK_2023} V. E. Adler, M. P. Kolesnikov, Non-autonomous reductions of the KdV equation and
multi-component analogs of the Painlevé equations P34 and P3, J. Math. Phys. 64:10 (2023) 101505.

\bibitem{Svinolupov_Sokolov_Yamilov} S. I. Svinolupov, V. V. Sokolov, R. I. Yamilov, On Bäcklund
transformations for integrable evolution equations, Dokl. Akad. Nauk SSSR 271:4 (1983) 802--805;
English transl.: Soviet Math. Dokl. 28 (1983) 165--168.

\bibitem{Garif_2023} R. N. Garifullin, Classification of semidiscrete equations of hyperbolic
type. The case of third-order symmetries, Theor. Math. Phys. 217:2 (2023) 1767--1776.

\bibitem{Garif_2021} R. N. Garifullin, I. T. Habibullin, Generalized symmetries and integrability
conditions for hyperbolic type semi-discrete equations, J. Phys. A: Math. Theor. 54 (2021) 205201,
doi:10.1088/1751-8121/abf3ea.

\bibitem{Adler_Veselov} V. E. Adler, A. P. Veselov, Cauchy problem for integrable discrete
equations on quad-graphs, Acta Appl. Math. 84 (2004) 237--262.

\bibitem{Atkinson_2012} J. Atkinson, A multidimensionally consistent version of Hirota's discrete
KdV equation, J. Phys. A: Math. Theor. 45 (2012) 222001.

\bibitem{KN_2012} P. Kassotakis, M. Nieszporski, On non-multiaffine consistent-around-the-cube
lattice equations, arXiv:1106.0435.

\bibitem{KN_2011} P. Kassotakis, M. Nieszporski, Families of integrable equations, SIGMA 7 (2011) 100.

\bibitem{KN_2018} P. Kassotakis, M. Nieszporski, Difference systems in bond and face variables and
non-potential versions of discrete integrable systems, arXiv:1710.11111.

\bibitem{ABS_YB} V. E. Adler, A. I. Bobenko, Yu. B. Suris, Geometry of Yang--Baxter maps: pencils
of conics and quadrirational mappings, Comm. Anal. Geom. 12 (2004) 967--1007.

\bibitem{PTV} V. G. Papageorgiou, A. G. Tongas, A. P. Veselov, Yang--Baxter maps and symmetries of
integrable equations on quad-graphs, J. Math. Phys. 47 (2006) 083502.

\bibitem{PSTV} V. G. Papageorgiou, Yu. B. Suris, A. G. Tongas, A. P. Veselov, On quadrirational
Yang--Baxter maps, SIGMA 6 (2010) 033.

\bibitem{Adler_1993} V. E. Adler, Recuttings of polygons, Funkts. Anal. Prilozhen. 27:2 (1993) 79--82; English transl.: Funct. Anal. Appl. 27:2 (1993) 141--143.

\bibitem{Atkinson_2013} J. Atkinson, Idempotent biquadratics, Yang--Baxter maps and birational
representations of Coxeter groups, arXiv:1301.4613.

\bibitem{Atkinson_2024} J. Atkinson, The tau function for ABS equations, arXiv:2410.21148.

\end{thebibliography}
\end{document}